\documentclass[12pt]{article}
\usepackage{latexsym}
\usepackage{amsmath}
\usepackage{amssymb}
\renewcommand{\thefootnote}{\fnsymbol{footnote}}
\newcommand{\newsection}{    
\setcounter{equation}{0}\section}
\def\appendix#1{\addtocounter{section}{1}\setcounter{equation}{0}
\renewcommand{\thesection}{\Alph{section}}
\section*{Appendix \thesection\protect\indent \parbox[t]{11.15cm}{#1}}
\addcontentsline{toc}{section}{Appendix \thesection\ \ \ #1}}

\def \bi{\bibitem}
\def \la {\label}

\def \b {\beta}

\def\det{\hbox{det}}
\def\be{\begin{equation}}
\def\ee{\end{equation}}

\def\iB{B}

\catcode`\@=11

\def\bea{\begin{eqnarray}}
\def\eea{\end{eqnarray}}
\def\beann{\begin{eqnarray*}}
\def\eeann{\end{eqnarray*}}
\def\beq{\begin{equation}}
\def\eeq{\end{equation}}
\def\ba{\begin{array}}
\def\ea{\end{array}}
\def\ben{\begin{enumerate}}
\def\een{\end{enumerate}}

 \def \la {\label}
 \def\be{\begin{equation}}
\def\ee{\end{equation}}

\def \la {\label}

\font\mybb=msbm10 at 11pt

\def\bb#1{\hbox{\mybb#1}}

\def\bZ {\bb{Z}}
\def\bR {\bb{R}}

\def\bC {\bb{C}}

\def\e  {\epsilon}

\def \ee {\epsilon}

\def \bi{\bibitem}
\def\a{\alpha }

\def \b {\beta}

\def\be{\begin{equation}}
\def\ee{\end{equation}}

\def \bi {\bibitem}
\def \la{\label}

\def \nn {\nonumber}

\begin{document}
\date{November 2002}
\begin{titlepage}
\begin{center}
\vspace*{-1.0cm}

\vspace{2.0cm} {\Large \bf Classification of IIB backgrounds with 28 supersymmetries} \\[.2cm]

\vspace{1.5cm}
 {\large  U. Gran$^1$, J. Gutowski$^2$ and  G. Papadopoulos$^2$}

\vspace{0.5cm}

${}^1$ Fundamental Physics\\
Chalmers University of Technology\\
SE-412 96 G\"oteborg, Sweden\\

\vspace{0.5cm}
${}^2$ Department of Mathematics\\
King's College London\\
Strand\\
London WC2R 2LS, UK\\

\vspace{0.5cm}

\end{center}

\vskip 1.5 cm
\begin{abstract}

We show that all IIB backgrounds with strictly  $28$ supersymmetries are locally isometric to the plane wave solution of arXiv:hep-th/0206195. Moreover,
we demonstrate that all solutions with more than 26 supersymmetries and only 5-form flux are maximally supersymmetric. The N=28 plane wave solution
is a superposition of the maximally supersymmetric IIB plane wave  with a heterotic string solution.
We investigate the propagation of strings in this background, find the spectrum  and give the string light-cone Hamiltonian.

\end{abstract}

\end{titlepage}
\newpage
\setcounter{page}{1}
\renewcommand{\thefootnote}{\arabic{footnote}}
\setcounter{footnote}{0}

\setcounter{section}{0}
\setcounter{subsection}{0}

\newsection{Introduction}

The geometry of backgrounds with a near maximal number of supersymmetries is strongly constrained. The maximally
supersymmetric IIB backgrounds have been classified in \cite{georgejose} and they have been found to be locally
isometric to Minkowski space, $AdS_5\times S^5$  \cite{schwarz} and the maximally supersymmetric plane wave \cite{plwav}.
It has also been shown that IIB backgrounds with more than 28 supersymmetries, $N>28$, are maximally supersymmetric \cite{n31, nm28},
and that IIB backgrounds with more than 24 supersymmetries are locally homogeneous \cite{hjose}. The latter implies
in particular that the 1-form field strength vanishes, $P=0$.
It is also known that there is a plane wave  solution in IIB supergravity with non-vanishing 3- and 5-form field strengths
which preserves 28 supersymmetries found by Bena and Roiban in \cite{benaroiban}, see also \cite{michelson}. So there are
IIB backgrounds with strictly 28 supersymmetries
which are not locally isometric to the maximally supersymmetric ones.

The main result of this paper is to show that all IIB supergravity backgrounds with strictly 28 supersymmetries are locally
isometric to the plane wave in  \cite{benaroiban}. This will be achieved using the spinorial geometry method of solving Killing spinor
equations (KSEs) \cite{mtheor} as
adapted to IIB supergravity in \cite{sgiib, sgiib2, iibsyst}  and to near maximally supersymmetric backgrounds in \cite{n31}.
In particular, the gauge symmetry
of IIB supergravity will be used to find the canonical form of the normals  to the 28 Killing spinors of the background.
Then the integrability condition of the gravitino KSE will be solved to reveal that the only solution
is that of \cite{benaroiban}. The proof is completed by showing that there are no N=28 IIB backgrounds
which can arise as discrete quotients of the maximally supersymmetric ones. This establishes the uniqueness of
\cite{benaroiban}, up to discrete identifications, as a IIB solution which preserves strictly 28 supersymmetries.

Another consequence of our analysis is that all $N>26$ IIB backgrounds with only 5-form flux are maximally supersymmetric.
This follows from the observation that if $G=0$, the $N=28$ backgrounds are maximally supersymmetric, and from the property that backgrounds with only 5-form flux preserve an even number of supersymmetries.

The main observation that allows our analysis to be carried out is that the Killing spinors of $N=28$ backgrounds can be expressed in terms
of a basis $(\eta_a, i\eta_a)$, where $(\eta_a)$ are 14 linearly independent spinors over the complex numbers. The algebraic
KSE can be easily solved by expressing the 3-form flux $G$ in terms of the normals to the Killing spinors.
Then the local part of the proof which involves the solution of the integrability conditions of the gravitino KSE is  separated into three
different cases labeled by the isotropy group of one of the normal spinors. These isotropy groups are $SU(4)\ltimes \bR^8$, $Spin(7)\ltimes \bR^8$
and $G_2$. In both the $SU(4)\ltimes \bR^8$ and $G_2$ cases, all backgrounds that admit $N=28$ supersymmetry are locally maximally
supersymmetric, and so they do not give new solutions. The solution of \cite{benaroiban} arises in the  $Spin(7)\ltimes \bR^8$ case.

The solution of \cite{benaroiban} can be interpreted as a superposition of the IIB maximally supersymmetric plane wave \cite{plwav}
with the solution of the heterotic string, see \cite{benaroiban} and also \cite{jose2},
which preserves 14 supersymmetries. The latter can be
``embedded'' into  IIB supergravity   and in such a case preserves 28 supersymmetries.
Using this interpretation, we investigate the
propagation of strings on this background. We find that the lightcone Hamiltonian is
the sum of harmonic oscillators and compute their
frequencies. We find that all directions of the center of mass mode of the string exhibit the
same frequency while the different directions of each oscillating mode
 exhibit two characteristic frequencies.

This paper is organized as follows. In section 2, we give the normals to the Killing spinors and solve the algebraic KSE. In section 3,
 we state the integrability conditions of the KSEs. In sections 4, 5 and 6, we solve the integrability conditions of the gravitino KSE
 in the $SU(4)\ltimes \bR^8$, $Spin(7)\ltimes \bR^8$ and $G_2$ cases, respectively. In section 7, we investigate the discrete quotients
 of maximally supersymmetric backgrounds. In section 8, we solve string theory on the $N=28$ plane wave background, and in section
 9 present our conclusions. In appendix A, we choose the normals to the Killing spinors up to gauge transformations. In appendix B,
 we summarize the integrability condition of the gravitino KSE, and in appendix C we present a part of the analysis for the $Spin(7)\ltimes \bR^8$
 case.

\newsection{The Algebraic Killing Spinor Equation}

\subsection{Normal spinors}

The main task here is to identify the four normals to the  Killing spinors of $N=28$ backgrounds.
In particular, we shall show that the normals are two spinors which are linearly independent over the complex numbers.
This in turn will imply that the Killing spinors can be expressed in terms of a basis $(\eta_a, i\eta_a)$,
where $\eta_a$ are 14 spinors linearly independent over the complex numbers.

{}For this consider the  algebraic KSE of IIB supergravity \cite{schwarz, west, howe}
\bea
{\cal A}\e\equiv P_A \Gamma^A C \epsilon^*+{1\over24} G_{ABC} \Gamma^{ABC}\e=0~,
\eea
where $P$ and $G$ are the 1-form and 3-form field strengths, respectively, and $A,B,C$ are spacetime frame indices.
Since all IIB backgrounds with more that 24 supersymmetries are homogeneous, the scalars are constant and $P=0$.
Therefore the algebraic KSE reduces to
\bea
G_{ABC} \Gamma^{ABC}\e=0~.
\eea
It is clear now that ${\cal A}$ is linear over the complex numbers,  ie if $\e$ is a solution so is $i\e$.

To continue suppose both algebraic and gravitino KSEs of a background admit 28 supersymmetries and let $\e_1, \dots, \e_{28}$ be the
 Killing spinors. It is required that $\e_1, \dots, \e_{28}$ are {\it linearly independent} over the {\it real} numbers. Since the algebraic
KSE is linear over the complex numbers, $i\e_1, \dots, i\e_{28}$ are also solutions of the algebraic equation. If one of these additional solutions
is linearly independent from $\e_1, \dots, \e_{28}$, over the reals, the dilatino KSE will admit more than 28
solutions. In such a case, we know that the only solution is $G=0$ \cite{nm28}. So there are two possibilities to consider for backgrounds
that preserve 28 supersymmetries.  Either the algebraic KSE admits a basis of $14$ linearly independent solutions, $\{\eta_a\}$, over the
complex numbers, and so Killing spinors can be written as
\bea
\e_r=\sum_{a=1}^{14} f_r{}^a\eta_a+\sum_{a=1}^{14} \tilde f_r{}^a i\eta_a~,~~~r=1, \dots, 28~,
\la{kspin}
\eea
where $(f, \tilde f)$ is an {\it invertible} real $28\times 28$ matrix of spacetime functions, or $G=0$.
In the $G=0$ case, the gravitino KSE also becomes linear over the complex numbers and so the Killing spinors
are in pairs $(\e_a, i\e_a)$. So the 28-plane of Killing spinors in both cases  is complex. Therefore, the Killing spinors
are {\it normal} to two spinors $\nu_1$ and $\nu_2$ which are {\it linearly
independent over the complex numbers}. We shall use the gauge symmetry of IIB supergravity to find canonical forms for
$\nu_1$ and $\nu_2$ and so simplify the choice of the basis $(\eta_a)$ of the Killing spinors.

\subsection{Solution to the algebraic KSE}

Assuming that $G\not=0$, the solution to the algebraic KSE can be expressed in terms of the normals to the Killing spinors.
Before we proceed to show this, we
take the Killing spinors to be in the positive chirality Weyl representation, $\Delta^+_{\bf 16}$, of $Spin(9,1)$. In such a case,
the normals to the Killing spinors, with respect to the Majorana inner product\footnote{We use the spinor conventions
of \cite{sgiib}. In particular,  $\iB(\theta, \zeta)=\langle B \theta^*, \zeta\rangle$, where $\langle \cdot, \cdot\rangle$ is a
Hermitian inner product and
 $B=\Gamma_{06789}$.}  $B$,
lie in the anti-chiral representation $\Delta^-_{\bf 16}$,  \cite{n31}.

To proceed, note
that there is an isomorphism $\Lambda^3 (\bR^{9,1}\otimes \bC) \cong
\Lambda^2 (\bR^{16}\otimes \bC)$ between the complexified 3-forms on
$\bR^{9,1}$ and the complexified 2-forms on $\bR^{16}$. In particular, identify $\bR^{16} \otimes \bC=\Delta^-_{\bf 16}$ and write
\bea
G_{A_1 A_2 A_3} = {1 \over 2}\lambda_{ij}\, \iB(\theta^i,\Gamma_{A_1 A_2 A_3}\theta^j)~,
\eea
where $\theta^i$ is a basis in $\Delta^-_{\bf 16}$.
Note that $B(\theta^i, \Gamma^{[3]}\theta^j)= - B(\theta^j, \Gamma^{[3]}\theta^i)$.

Next we have the identity
\be
{1 \over 3!} \iB (\phi_1, \Gamma_{A_1 A_2 A_3} \phi_2) \Gamma^{A_1 A_2 A_3} \phi_3
= 8  \iB (\phi_2, \phi_3 )  \phi_1 -8 \iB(\phi_1, \phi_3) \phi_2~,
\ee
where  $\phi_1, \phi_2 \in \Delta^-_{\bf 16}$ and $\phi_3 \in \Delta^+_{\bf 16}$. Applying this to the
algebraic KSE, we have
\be
\label{fierz1}
{1 \over 3!} G_{A_1 A_2 A_3} \Gamma^{A_1
A_2 A_3} \eta_a = 8 \lambda_{ij}
\iB ( \theta^j, \eta_a )  \theta^i~.
\ee
Choosing $\theta^a=B\eta_a^*$, $a=1,\dots,14$, $\theta^{15}=\nu_1$ and $\theta^{16}=\nu_2$, we have that the above equation vanishes
iff
\be
\lambda_{ab}=0, \qquad \lambda_{15, a} = \lambda_{16, a}=0 \qquad a,b=1,
\dots , 14~.
\ee

It follows that the solution of the algebraic KSE is
\be
\label{gbilin}
G_{A_1 A_2 A_3} = \lambda\,
\iB (\nu_1, \Gamma_{A_1 A_2 A_3}
\nu_2 )~
\ee
for $\lambda$ a complex function.  Since in the spinorial geometry approach
the normal spinors are
determined up to gauge transformations, (\ref{gbilin}) can be used to compute $G$.
If $G \neq 0$, then after rescaling one of the normal spinors we can set $\lambda =1$. For future use observe that
\be
\nabla_{A_1} G_{A_2 A_3 A_4} = \iB (\nabla_{A_1} \nu_1, \Gamma_{A_2 A_3 A_4}
\nu_2 ) + \iB (\nu_1, \Gamma_{A_2 A_3 A_4}
\nabla_{A_1} \nu_2 )~,
\ee
where $\nabla$ is the frame Levi-Civita connection. We shall show that for all $N=28$ backgrounds, $G$ is parallel.

\newsection{Integrability Conditions}

To make further progress, we shall investigate the integrability conditions of the
 KSEs
\bea
{\cal D}_M\e&\equiv& \nabla_M\e+{i\over48} \Gamma^{N_1\dots N_4} F_{N_1\dots N_4 M}\e-
{1\over 96} (\Gamma_M{}^{N_1N_2N_3}G_{N_1N_2N_3}-9 \Gamma^{N_1N_2} G_{MN_1N_2})C\e^*=0~,
\cr
{\cal A}\e&\equiv&  G_{M_1 M_2 M_3} \Gamma^{M_1 M_2 M_3} \epsilon=0~,
\eea
where we have set $P=0$ as we have already explained. Since the matrix $(f, \tilde f)$ in (\ref{kspin}) is invertible, the integrability conditions
on the Killing spinors can be evaluated on the basis $(\eta_a, i\eta_a)$. Because of the complex nature of this basis, as we shall demonstrate,
the integrability conditions factorize.

{}First, we take the $\nabla$-derivative of the algebraic KSE and then substitute for $\nabla\e$ using the gravitino KSE to find
\bea
\label{mixedint}
\big(\nabla_M G_{N_1 N_2 N_3} \Gamma^{N_1 N_2 N_3}
-{i \over 2} G_{N_1 N_2 L} F_M{}^L{}_{N_3 N_4 N_5}
\Gamma^{N_1 N_2 N_3 N_4 N_5} + i G_{N_1 N_2 N_3}
F_M{}^{N_1 N_2 N_3}{}_L \Gamma^L \big) \epsilon
\nn \\
-{1 \over 96} \big[\Gamma_M (G_{N_1 N_2 N_3} \Gamma^{N_1 N_2 N_3})
 (G_{N_4 N_5 N_6} \Gamma^{N_4 N_5 N_6})
 +6 G_{M L_1 L_2} \Gamma^{L_1 L_2}  (G_{N_1 N_2 N_3} \Gamma^{N_1 N_2 N_3})
 \nn \\
 +144 G_{N_1 N_2 L} G_{N_3 M}{}^L \Gamma^{N_1 N_2 N_3} \big] C \epsilon^* =0~.
 \nn \\
 \eea

Evaluating this condition on the Killing spinor (\ref{kspin})  basis $(\eta_a, i\eta_a)$, observe that it factorizes as
\bea
\label{factorint1}
\big(\nabla_M G_{N_1 N_2 N_3} \Gamma^{N_1 N_2 N_3}
-{i \over 2} G_{N_1 N_2 L} F_M{}^L{}_{N_3 N_4 N_5}
\Gamma^{N_1 N_2 N_3 N_4 N_5}
\nn \\
 + i G_{N_1 N_2 N_3}
F_M{}^{N_1 N_2 N_3}{}_L \Gamma^L \big) \eta_a =0~,
\eea
and
\bea
\label{factorint2}
 \big[\Gamma_M (G_{N_1 N_2 N_3} \Gamma^{N_1 N_2 N_3})
 (G_{N_4 N_5 N_6} \Gamma^{N_4 N_5 N_6})
 +6 G_{M L_1 L_2} \Gamma^{L_1 L_2}  (G_{N_1 N_2 N_3} \Gamma^{N_1 N_2 N_3})
 \nn \\
 +144 G_{N_1 N_2 L} G_{N_3 M}{}^L \Gamma^{N_1 N_2 N_3} \big] C \eta_a^* =0~,
 \nn \\
\eea
for $a=1, \dots, 14$.

In addition, the gravitino KSE integrability condition,
\bea
[{\cal D}_N, {\cal D}_M]\e\equiv {\cal R}_{NM}\e= 2{\cal S}\e-2{\cal T}C\e^*~,
\eea
 implies that
\be
\label{grav1}
{\cal{S}} \eta_a =0~,
\ee
and
\be
\label{grav2}
{\cal{T}}C \eta_a^* =0~,
\ee
where ${\cal S}$ and ${\cal T}$ are given in  \cite{tsimpis} and the special case $P=0$ that applies here is stated in appendix B for convenience.

In what follows, we shall investigate the above integrability  conditions for the various choices of Killing spinors which are specified by choosing
their normals up to gauge transformations. It is convenient to
label the various cases with the isotropy group of the first normal in the $Spin(9,1)$ gauge group.

\newsection{$SU(4) \ltimes \bR^8$}
\label{su4}

 \subsection{Normal spinors}

 A representative for the first $SU(4)\ltimes\bR^8$-invariant normal \cite{n31, nm28} is
 \be
\nu_1 = -p e_5 - q e_{12345}~,
\ee
where $p, q$ are complex functions with $|p| \neq |q|$. Observe that if $|p|=|q|$, then $\nu_1$ is $Spin(7)\ltimes\bR^8$-invariant and this case
will be examined separately. To choose the second normal $\nu_2$ in the $SU(4)\ltimes\bR^8$ case, one has to decompose
$\Delta_{\bf 16}^-$ under $SU(4)\ltimes\bR^8$ and choose representatives for the various orbits, see appendix A.
As is mentioned in appendix A the choice of the second normal can be simplified by assuming that the 1-form bilinear of any linear
combination of $\nu_1$ and $\nu_2$ is null. This is because if a direction in the $(\nu_1, \nu_2)$-plane is associated with a time-like
1-form bilinear, then the corresponding solutions are special cases of  $G_2$ backgrounds we shall analyze in section \ref{G2}.

To summarize the detailed analysis in appendix A, there are three choices for the normals. These are
\be
 \label{norm2ax2b}
 \nu_1 = - p e_5 -q e_{12345} , \qquad \nu_2 = -y e_{12345} - u^1 e_1 - w e_{234}, \
\ee
with $\bar{w} p + u^1 \bar{q} =0$ and  $p,w,q, u^1\not=0$, where $\bar w$ is the complex conjugate of $w$ and similarly for the other functions, or
\be
 \label{norm2ax1b}
\nu_1 = e_5, \qquad \nu_2 = c\,\, e^1 \qquad (c \neq 0)~,
\ee
or
\be
\label{normal2bx}
\nu_1 = - p e_5 -q e_{12345} , \qquad \nu_2 = -x e_5 -y e_{12345} -c_1 e_{145} -c_2 e_{235}~.
\ee

\subsection{Solutions with $G \neq 0$}

Here we shall solve the integrability conditions for all the three choices of normals.

\subsubsection{$\nu_1 =  p e_5 -q e_{12345}$ ,  $\nu_2 = -y e_{12345} - u^1 e_1 - w e_{234}$}

This choice of normal leads to a basis $(\eta_a)$ in the space of Killing spinors which includes the spinors
\bea
\{ e_{1235}, e_{1245}, e_{25}, e_{35}, e_{45}, e_{13}, e_{23}, e_{24}, e_{34}, e_{1345} \}~.
\eea
Substituting each of these spinors in the integrability condition ({\ref{factorint2}}) and assuming that $G\not=0$,
one finds that $w=0$. This is a contradiction because for this choice of normals $w \neq 0$. Hence, there are no solutions unless $G=0$
which will be considered separately.

\subsubsection{$\nu_1 = e_5$, $\nu_2 = c\, e^1$ }

{}For this choice of normals, ({\ref{factorint2}}) is automatically satisfied.  To proceed further, consider applying ({\ref{factorint1}})
and ({\ref{grav2}}) to the spinors orthogonal
to $\nu_1 = e_5,  \nu_2 = c e_{1} $.
These integrability conditions imply
 \bea
 F=0~.
 \eea
On the other hand using (\ref{gbilin}), the 3-form $G$ can be written as
\bea
G &=& c \big( -e^2\wedge e^3\wedge e^4+e^2\wedge e^8\wedge e^9-e^3\wedge e^7\wedge e^9+e^4\wedge e^7\wedge e^8
\cr
&&~~~~~
-ie^2\wedge e^3\wedge e^9+ie^2\wedge e^4\wedge e^8-ie^3\wedge e^4\wedge e^7+ie^7\wedge e^8\wedge e^9 \big)~,
\eea
and so
\be
G \wedge G^* = 8i |c|^2\, e^2\wedge e^3\wedge e^4\wedge e^7\wedge e^8\wedge e^9
\la{gg}
\ee
which does not vanish for $c \neq 0$. Since $F=0$, these data are incompatible with the Bianchi identity of $F$ for which $dF$ is proportional to
(\ref{gg}).
Hence, there are no solutions unless $G=0$
which will be considered separately.

\subsubsection{$\nu_1 = -p e_5 - q e_{12345}$,  $\nu_2 = -x e_5 -y e_{12345} -c_1 e_{145} -c_2 e_{235}$}

There are a number of cases to consider.
First, if $c_1=c_2=0$ and insisting that $\nu_1$ and $\nu_2$ are linearly independent, then a basis in the $(\nu_1, \nu_2)$-plane can be chosen
such that the first normal spinor is $Spin(7) \ltimes \bR^8$ invariant. Therefore this is a special case of  backgrounds with a
$Spin(7) \ltimes \bR^8$-invariant normal which will be examined  separately.

Second, if one of $c_1$ or $c_2$ does not vanish, without loss of generality,
one can take $c_1 \neq 0$. By applying a $SU(4)$ transformation, we can
take ${c_2 \over c_1}$ to be a {\it real} function. In such a case, a basis $(\eta_a)$ in the space of Killing spinors
can be chosen to include the 13 spinors
\be
\{ e_{15}, e_{25}, e_{35}, e_{45}, e_{12}, e_{13}, e_{24}, e_{34}, e_{1235}, e_{1245}, e_{1345},
e_{2345}, {c_2 \over c_1}e_{23}-e_{14} \}~.
\ee
Substituting this into ({\ref{factorint2}}),
we find the relations
\be
pq (c_2^2-c_1^2)=0 , \quad q(yp-xq) (c_2^2-c_1^2)=0, \quad p(yp-xq) (c_2^2-c_1^2)=0~.
\ee

The solution of the above relations leads to  three further sub-cases:

\begin{itemize}

\item[(i)] $c_2 \neq \pm c_1$, $p=x=0$ and $q \neq 0$, which gives
\be
\nu_1 = e_{12345}, \quad \nu_2 = -c_1 e_{145} - c_2 e_{235}
\ee

\item[(ii)] $c_2 \neq \pm c_1$, $q=y=0$ and $p \neq 0$, which gives
\be
\nu_1 = e_{5}, \quad \nu_2 = -c_1 e_{145} - c_2 e_{235}
\ee

\item[(iii)] $c_2 = \pm c_1$. After a $SU(4)$ transformation to set $c_1=c_2$ and then re-scaling of $\nu_2$, one finds
\bea
\nu_1 = -p e_5 -q e_{12345}, \quad \nu_2 = -x e_5 -y e_{12345} -e_{145} -e_{235}
\eea

\end{itemize}

Further simplification of the above three cases is possible  by applying ({\ref{factorint2}})
to the 14-th basis element
\be
\eta_{14} = p1 -q e_{1234} + {(py-qx) \over c_1} e_{23}
\ee
in the space of Killing spinors.

In particular, for the $c_2=c_1=1$ case, one obtains
\be
(|p|^2-|q|^2)(2|p|^2+2|q|^2+|yp-xq|^2)=0~.
\ee
Therefore, $|p|= |q|$, and thus this solution is a special case of those for which  $\nu_1$ is $Spin(7) \ltimes \bR^8$ invariant.

{}For the other two cases for which  $c_1 \neq \pm c_2$,   ({\ref{factorint2}}) evaluated on $\eta_{14}$ implies that   $c_2=0$. By using the gauge transformation $e^{{\pi \over 2} (\Gamma_{12}+\Gamma_{34})}$ (with real basis indices),
together with
appropriately chosen $SU(4)$ gauge transformations, one can simplify the normals as
\be
\nu_1 = e_5, \qquad \nu_2 = c\, e_{345} \ .
\la{nnn}
\ee
To summarize so far, after solving ({\ref{factorint2}}), the only choice of normals allowed provided one of them is $SU(4)\ltimes\bR^8$-invariant
is given in (\ref{nnn}).

Next let us turn to investigate the remaining integrability conditions for the Killing spinors normal to (\ref{nnn}).
 To proceed, we use (\ref{gbilin}) to write $G$ as
\be
G = \sqrt{2}\, c\, e^+ \wedge \big( e^1\wedge e^2-e^6\wedge e^7+ie^1\wedge e^7-ie^2\wedge e^6\big)~.
\la{gggg}
\ee
 A basis $(\eta_a)$ for the Killing spinors
normal to (\ref{nnn}) is
\bea
 \{ 1, e_{15}, e_{25}, e_{35}, e_{45}, e_{13}, e_{14}, e_{23},
e_{24}, e_{34}, e_{1235}, e_{1245}, e_{1345}, e_{2345} \}~.
\eea
Then
 ({\ref{factorint1}})
and ({\ref{grav2}}) imply that
\bea
F=0~,~~~\nabla G=0~,
\eea
ie $G$ is parallel with respect to the Levi-Civita connection.

It remains to solve the integrability condition  ${\cal{S}} \eta_a =0$. Since $G$ is null (\ref{gggg}), the terms $G$-quadratic terms in ${\cal S}$  can be simplified to write
\bea
{\cal{S}}_{NM} &=& {1 \over 8} R_{NM,L_1 L_2} \Gamma^{L_1 L_2}
+{1 \over 32} \big(-{1 \over 8} \Gamma^{L_1 L_2} G_{NM}{}^Q G^*_{L_1 L_2 Q}
- {13 \over 12} \Gamma^{L_1 L_2}  G^*_{NM}{}^Q G_{L_1 L_2 Q}
\nn \\
&-&{1 \over 48} \Gamma^{L_1 L_2 L_3 L_4} G_{NML_1} G^*_{L_2 L_3 L_4}
+{1 \over 4} \Gamma_{[N|}{}^{L_1 L_2 L_3} G_{|M] L_1}{}^Q G^*_{L_2 L_3 Q}
\big)~.
\eea

To proceed, if $M= {\tilde{M}}$, $N={\tilde{N}}$, where
${\tilde{M}}$, ${\tilde{N}}$ take all values except for ``+", and if  $N=+$, $M=-$, then we obtain
\be
 R_{ {\tilde{N}} {\tilde{M}}, L_1 L_2} \Gamma^{L_1 L_2} \eta_a =0~,~~~ R_{ + -, L_1 L_2} \Gamma^{L_1 L_2} \eta_a =0~.
\ee
On applying $C*$ to both these conditions, we find that
\be
  R_{ {\tilde{N}} {\tilde{M}}, L_1 L_2} \Gamma^{L_1 L_2} \eta =0,
 \qquad
  R_{ + -, L_1 L_2} \Gamma^{L_1 L_2} \eta =0
 \ee
for all Majorana-Weyl spinors $\eta$, which in turn implies that the associated Riemann curvature components vanish. Hence the only non-vanishing components
of the Riemann tensor are $R_{+i+j}$ for $i,j=1,2,3,4,6,7,8,9$.

To continue, it is convenient to rewrite the remaining ${\cal S}$  integrability conditions as
\be
\big( {1 \over 2} (T^2_{MN})_{L_1 L_2} \Gamma^{L_1 L_2}
+{1 \over 24} (T^4_{MN})_{L_1 L_2 L_3 L_4} \Gamma^{L_1 L_2 L_3 L_4} \big) \eta_a =0~,
\la{st2t4}
\ee
where
\be
(T^2_{MN})_{L_1 L_2} = R_{NM,L_1 L_2}-{1 \over 32}G_{NM}{}^Q G^*_{L_1 L_2 Q}
-{15 \over 64} G^*_{NM}{}^Q G_{L_1 L_2 Q}~,
\ee
and
\be
(T^4_{MN})_{L_1 L_2 L_3 L_4} = -{1 \over 16}G_{NM [L_1}G^*_{L_2 L_3 L_4]}
+{3 \over 8} \delta_{N[L_1} G_{|M| L_2}{}^Q G^*_{L_3 L_4] Q}
- {3 \over 8} \delta_{M[L_1} G_{|N| L_2}{}^Q G^*_{L_3 L_4] Q}~.
\ee
It is straightforward to show that the only non-vanishing components
of $T^2$ and $T^4$ are $(T^2_{+i})_{+j}$, $(T^4_{+i})_{+ q_1 q_2 q_3}$.
Using this, the integrability condition  ${\cal{S}} \eta_a =0$ is equivalent to
\be
\big( (T^2_{+i})_{+j} \Gamma^j +{1 \over 6} (T^4_{+i})_{+ j_1 j_2 j_3}
\Gamma^{j_1 j_2 j_3} \big) \chi_a =0
\la{t2t4}
\ee
for $\chi_a \in \{ e_5, e_{135}, e_{145}, e_{235}, e_{245}, e_{345} \}$.
In order to analyse the conditions imposed by these integrability conditons we
have used a computer assisted computation (CAC)\footnote{We can provide more information on request.}.
One finds that $c=0$, however this is a contradiction, since we have assumed $G \neq 0$.
In conclusion, in all cases, we deduce that  we should take $G=0$.

\subsection{Solutions with $G=0$}

To investigate  the solutions with $G=0$, we write the gravitino integrability condition as
\be
\label{maxint1}
{\cal S}\eta_a\equiv\bigg( {1 \over 2} (T^2_{MN})_{N_1 N_2} \Gamma^{N_1 N_2} +{1 \over 4!} (T^4_{MN})_{N_1 N_2 N_3 N_4}
\Gamma^{N_1 N_2 N_3 N_4} \bigg) \eta_a =0
\ee
where now
\bea
(T^2_{MN})_{P_1P_2} &=& \tfrac{1}{4} R_{M N, P_1P_2}-\tfrac{1}{12}F_{M[P_1}{}^{Q_1Q_2Q_3}F_{|N|P_2]Q_1Q_2Q_3}~,\notag\\
(T^4_{MN})_{P_1 \ldots P_4} &=& \tfrac{i}{2}D_{[M}F_{N]P_1\ldots P_4}+\tfrac{1}{2}F_{M N Q_1Q_2 [P_1}F_{P_2 P_3 P_4]}{}^{Q_1 Q_2}~.
\label{Tphys}
\eea
The field equations and Bianchi identities imply that
\bea
(T^2_{MN})_{P_1P_2} &=& (T^2_{P_1P_2})_{MN}~,~~~(T^2_{M[P_1})_{P_2P_3]} = (T^2_{MN})_{P}{}^N = 0~,
\cr
(T^4_{[P_1P_2})_{P_3 P_4 P_5 P_6]} &=&(T^4_{MN})_{P_1P_2P_3}{}^N = 0 ~,~~(T^4{}_{P_1 (M})_{N) P_2 P_3 P_4}=(T^4{}_{[P_1 |(M})_{N)| P_2 P_3 P_4]}~,
\cr
(T^4_{M[P_1})_{P_2 P_3 P_4 P_5]} &=& -{1 \over 5!}
\epsilon_{P_1 P_2 P_3 P_4 P_5}{}^{Q_1 Q_2 Q_3 Q_4 Q_5}
(T^4_{M[Q_1})_{Q_2 Q_3 Q_4 Q_5]}~.
\label{feqcond}
\eea

In the $SU(4)\ltimes\bR^8$ case, there are two choices of normal spinors that we should consider up to $Spin(9,1)$ transformations.
{}First consider the case in which the two normals can be chosen as
\be
\nu_1 = -p e_5 -q e_{12345}, \qquad \nu_2 = -y e_{12345} - u^1 e_1 -u^2 e_2 -w e_{234} -c_3 e_{235}~,
- c_4 e_{345}~,
\la{nu1nu2}
\ee
and $|u^1|^2 + |u^2|^2 \neq 0$. This case can be further separated, as in the analysis of the
previous section,  into two different sub-cases
using the additional condition that the associated 1-form
bi-linears of all directions in the $(\nu_1, \nu_2)$-plane are null, see appendix A. However, there is no advantage to do this here
and so we shall treat both sub-cases together.
The basis $(\eta_a)$ of Killing spinors normal to $(\nu_1, \nu_2)$ in (\ref{nu1nu2}) includes the  spinors
\be
\{ e_{1235}, e_{1245}, e_{25}, e_{35}, e_{45}, e_{13}, e_{23}, e_{24}, e_{34},
w e_{1345} + u^2 e_{15}, w e_{2345} - u^1 e_{15} \}~.
\ee
Substituting these 11 spinors into ({\ref{maxint1}}) and making use of (\ref{feqcond}), one obtains $T^2=T^4=0$, so these solutions are
locally maximally supersymmetric. Here and in two similar cases below, we have again used CAC.

Next, consider the  case for which
\be
\nu_1 = -p e_5 -q e_{12345}, \qquad \nu_2 = -x e_5 -y e_{12345} -c_1 e_{145} - c_2 e_{235}~.
\la{2k2}
\ee
To proceed further, it is convenient to in addition assume that  $|c_1|^2+|c_2|^2 \neq 0$. In such a case, the basis $(\eta_a)$
of Killing spinors includes
\be
\{ e_{15}, e_{25}, e_{35}, e_{45}, e_{12}, e_{13}, e_{24}, e_{34},
e_{1235}, e_{1245}, e_{1345}, e_{2345}, c_2 e_{23} - c_1 e_{14} \}~.
\ee
Substituting these 13 spinors in ({\ref{maxint1}})  and using (\ref{feqcond}), one finds $T^2=T^4=0$.
  Again, these solutions are locally maximally supersymmetric.
It remains to consider the case for which $c_1=c_2=0$ in (\ref{2k2}). Then a basis in the space of Killing spinors is
\be
(\eta_a)=\{ e_{15}, e_{25}, e_{35}, e_{45}, e_{12}, e_{13}, e_{24}, e_{34},
e_{1235}, e_{1245}, e_{1345}, e_{2345}, e_{23}, e_{14} \}~.
\ee
Substituting this basis into ({\ref{maxint1}}) and using (\ref{feqcond}), one  finds that  $T^2=T^4=0$. So the solutions
are again locally maximally supersymmetric.

To summarize, if one of the two normal spinors of backgrounds preserving $N=28$ supersymmetries is $SU(4)\ltimes\bR^8$-invariant, then
they are locally maximally supersymmetric. Later we shall show that there are no quotients
of maximally supersymmetric backgrounds preserving 28 supersymmetries. As a consequence, all such $N=28$ supersymmetric backgrounds
 are maximally supersymmetric.

\newsection{$Spin(7) \ltimes \bR^8$-invariant normal}

\subsection{Solutions with $G \neq 0$}

It is explained in appendix A  that the two normal  spinors can be chosen as
\be
\nu_1 =e_5+ e_{12345}, \quad \nu_2 = c(e_5-e_{12345}), \qquad (c \neq 0)~.
\la{nspin7}
\ee
Using this and (\ref{gbilin}), one finds that
\be
G =2 \sqrt{2}\, i\, c\, e^+ \wedge \omega~,~~~~\omega= e^1\wedge e^6+e^2\wedge e^7+e^3\wedge e^8+ e^4\wedge e^9~.
\la{gspin7}
\ee

To proceed, a basis in the space of Killing spinors normal to $(\nu_1, \nu_2)$ given in (\ref{nspin7}) is
\bea
(\eta_a)=\{ e_{15}, e_{25}, e_{35}, e_{45}, e_{1235}, e_{1245}, e_{1345}, e_{2345},
e_{12}, e_{13}, e_{14}, e_{23}, e_{24}, e_{34} \}~.
\la{kspin7}
\eea
Substituting this into the integrability conditions
 ({\ref{factorint1}})
and ({\ref{grav2}}) and after some CAC, one finds that
\bea
\nabla (c \, e^+)=0~,~~~\nabla G=0~,~~~F=  f\, e^+\wedge \omega \wedge \Psi~,~~~
\la{fspin7}
\eea
where $f=|c|$ and
 $\Psi$ is a (1,1)- and $\omega$-traceless form in the directions $12346789$ transverse to the light-cone, ie
 \bea
\Psi_{kl}\, \omega^k{}_i \omega^l{}_j=\Psi_{ij}~,~~~\Psi_{ij} \omega^{ij}=0~.
 \eea
 Thus $c \,e^+$ and $G$ are {\it $\nabla$-parallel}.
In particular, as  $({\rm Re \ }c)  e^+$
and $({\rm Im \ } c) e^+$ are both covariantly constant, this implies that
there exists a {\it constant angle} $\phi$ such that $c = f e^{i \phi}$. Hence, the spacetime admits  a covariantly constant real 1-form
$V= f e^+$. Thus,
the spacetime geometry is that of a pp-wave.

It remains to evaluate the last integrability condition ${\cal{S}} \eta_a=0$, (\ref{grav1}), on the basis
(\ref{kspin7}) of the Killing spinors.  The expression for ${\cal{S}}$
can be considerably simplified  by making use of the special form for $F$ and $G$ which we have
obtained in (\ref{fspin7}) and (\ref{gspin7}), respectively. In particular, one can write

\bea
{\cal{S}} \eta_a\equiv \big( {1 \over 8} R_{NM,L_1  L_2} \Gamma^{L_1 L_2} -{i \over 48} \Gamma^{L_1 L_2 L_3 L_4}
D_{[N} F_{M] L_1 L_2 L_3 L_4}
-{1 \over 24} \Gamma^{L_1 L_2} F_{[N|L_1}{}^{Q_1 Q_2 Q_3} F_{|M] L_2 Q_1 Q_2 Q_3} \nn \\
+{1 \over 48} \Gamma^{L_1 L_2 L_3 L_4} F_{NM L_1}{}^{Q_1 Q_2} F_{L_2 L_3 L_4 Q_1 Q_2}
-{1 \over 32} \Gamma^{L_1 L_2} G_{NM}{}^{L_3} G^*_{L_1 L_2 L_3} \big) \eta_a =0~. \nn \\
\eea

Next observe that the 14-plane spanned by the basis (\ref{kspin7}) of the Killing spinors is invariant under the reality operation  $C*$,
ie $C*\{\eta_a\}=\{\eta_a\}$. Moreover, using that  $G=i f e^{i\phi} H$, where
$H$ is a {\it real} 3-form, which in turn implies that the $G$-quadratic terms in ${\cal S}$ are real, one finds that ${\cal{S}} \eta_a=0$ factorizes as
\be
\label{simpspin7int1}
\Gamma^{L_1 L_2 L_3 L_4}
D_{[N} F_{M] L_1 L_2 L_3 L_4} \eta_a =0~,
\ee
and
\bea
\label{simpspin7int2}
\bigg( \big( {1 \over 2} R_{NM, L_1 L_2} - {1 \over 8} G_{NM}{}^Q G^*_{L_1 L_2 Q}
-{1 \over 6} F_{[N|L_1}{}^{Q_1 Q_2 Q_3} F_{|M] L_2 Q_1 Q_2 Q_3} \big) \Gamma^{L_1 L_2} \nn \\
+{1 \over 12}   F_{NM L_1}{}^{Q_1 Q_2} F_{L_2 L_3 L_4 Q_1 Q_2}  \Gamma^{L_1 L_2 L_3 L_4} \bigg) \eta_a
=0~.
\eea

Let us first focus on ({\ref{simpspin7int2}}). Setting   $N={\hat{N}}$ and $M={\hat{M}}$, where ${\hat{N}}$ and ${\hat{M}}$ take all
values apart from ``+", and $N=+$ and $M=-$, and using the fact that both $F$ and $G$ are null, one finds that
\bea
R_{{\hat{N}} {\hat{M}}, L_1 L_2}  \Gamma^{L_1 L_2} \eta_a=R_{+ -, L_1 L_2}  \Gamma^{L_1 L_2} \eta_a =0~.
\eea
Since the isotropy group of $14$ linearly independent spinors in $Spin(9,1)$ is $\{1\}$, one concludes
that
\bea
R_{{\hat{N}} {\hat{M}}, L_1 L_2}=R_{+ -, L_1 L_2}=0~,
\eea
and so the only non-vanishing components of the Riemann tensor are
 $R_{+i,+j}$.

To proceed further, it is useful to define
\bea
(T^2_{NM})_{L_1 L_2} &=& R_{NM, L_1 L_2}-{1 \over 4} G_{NM}{}^Q G^*_{L_1 L_2 Q}
-{1 \over 3} F_{[N| L_1}{}^{Q_1 Q_2 Q_3} F_{|M] L_2 Q_1 Q_2 Q_3}~,
\nn \\
(T^4_{NM})_{L_1 L_2 L_3 L_4} &=& 2 F_{NM[L_1}{}^{Q_1 Q_2} F_{L_2 L_3 L_4] Q_1 Q_2}~.
\eea
In which case, ({\ref{simpspin7int2}}) can be rewritten as
\be
\big( {1 \over 2}(T^2_{NM})_{L_1 L_2} \Gamma^{L_1 L_2}
+{1 \over 24} (T^4_{NM})_{L_1 L_2 L_3 L_4} \Gamma^{L_1 L_2 L_3 L_4}
\big) \eta_a =0~.
\ee
As the only nonzero components of $T^2$ and $T^4$ are
$(T^2_{+i})_{+j}$ and $(T^2_{+i})_{+ \ell_1 \ell_2 \ell_3}$, respectively,
the only non-identically vanishing components of the above equation are
\be
\big( (T^2_{+i})_{+j} \Gamma^j +{1 \over 6} (T^4_{+i})_{+ \ell_1 \ell_2 \ell_3}
\Gamma^{\ell_1 \ell_2 \ell_3} \big) \Gamma_- \eta_a =0~,
\ee
or equivalently
\be
\big( (T^2_{+i})_{+j} \Gamma^j +{1 \over 6} (T^4_{+i})_{+ \ell_1 \ell_2 \ell_3}
\Gamma^{\ell_1 \ell_2 \ell_3} \big) \chi_a =0~,
\ee
where $\chi_a \in \{ e_{125}, e_{135}, e_{145}, e_{235}, e_{245}, e_{345} \}$.
It is straightforward to analyse these conditions,
and one finds that
\bea
T^2=T^4=0~.
\eea

In particular, $T^4=0$  implies that
\be
\label{algpsi}
\Psi_{\alpha}{}^\nu \Psi_{\nu \bar{\beta}} + {1 \over 4}\delta_{\alpha \bar{\beta}}
\Psi^{\nu \bar{\sigma}} \Psi_{\nu \bar{\sigma}} =0 \ ,
\ee
where the holomorphic indices $\a, \b, \nu, \mu=1,2,3,4$ are taken with respect to $\omega$.
The condition $T^2=0$ expresses the Riemann tensor of the spacetime in terms of the fluxes and we shall return  to it later.

It remains to solve the integrability condition  ({\ref{simpspin7int1}}). This is done in appendix C to find that
\bea
\nabla F=0
\eea
ie $F$ is also covariantly constant.

Returning to the condition $T^2=0$, since both $F$ and $G$ are covariantly constant, one concludes that
\bea
\nabla R=0
\eea
ie the spacetime is a Lorentzian symmetric space. These have been classified in \cite{cw}. Since in addition the spacetime
admits a $\nabla$-parallel null vector field
and the only non-vanishing components of the curvature are
\be
\label{riemconaux1}
R_{+i,+j} = \big( 2 f^2 + {1 \over 2} f^2 \Psi_{kl} \Psi^{kl} \big) \delta_{ij}~,
\ee
the spacetime is a {\it plane wave} or equivalently a Cahen-Wallach space. The above components of the Riemann tensor
determine the wave profile.

To find the background explicitly, since the fluxes and the Riemann curvature are covariantly constant, one can follow the analysis
of \cite{georgejose}  for the maximally supersymmetric plane wave. In particular, one can determine the fluxes at the origin
of the symmetric space. Then they can be defined everywhere on spacetime by acting with the transitive group.
Indeed, the expression for the spacetime geometry can be simplified somewhat by solving
(\ref{algpsi}). Since $\Psi$ is (1,1) and $\omega$-traceless, it lies up to a $SU(4)$-transformation on the maximal torus
of $\mathfrak{su}(4)$. Using this and (\ref{algpsi}), one finds that, without loss of generality, $\Psi$ can be written as
\bea
\Psi=-h (e^1\wedge e^6+ e^2\wedge e^7-e^3\wedge e^8-e^4\wedge e^9)~,
\eea
where $h$ is a real constant.
Adapting coordinates to the null vector field $f e^+=dv$ and putting the plane wave in Brinkman coordinates, the solution can be written as
\bea
ds^2 &=& 2 dv [du - (\ell^2+2h^2) \delta_{ij}x^i x^j dv] + \delta_{ij} dx^i dx^j
\nn \\
G &=&-2 \sqrt{2} i\, \ell\, e^{i \phi} dv \wedge (dx^1 \wedge dx^6 + dx^2 \wedge dx^7
+ dx^3 \wedge dx^8 + dx^4 \wedge dx^9)~,
\nn \\
F &=& 2h dv \wedge (dx^1 \wedge dx^2 \wedge dx^6 \wedge dx^7
-dx^3 \wedge dx^4 \wedge dx^8 \wedge dx^9)~,
\la{bena}
\eea
where we have re-instated a constant parameter $\ell$ using a coordinate transformation $v\rightarrow \ell^{-1} v$, $u\rightarrow \ell u$
and redefining the parameter as $h\rightarrow \ell h$. In the form given in (\ref{bena}) the solution depends on two parameters $(\ell, h)$
though one of them can be removed using a coordinate transformation provided that $\ell, h\not=0$. However, the form given in (\ref{bena})
allows us to also consider the limits in which these parameters vanish.
If either $h=0$ or $\ell=0$, the solution corresponds to either the heterotic solution of \cite{jose2}, $G$ real,  which preserves
14 supersymmetries or to the maximally supersymmetric plane wave solution of \cite{plwav} respectively. If both $F$ and $G$ are non-vanishing,
the solution preserves strictly 28 supersymmetries in IIB and it has been found in
\cite{benaroiban}.

\subsection{Solutions with $G=0$}

These backgrounds are a special case of the
 $SU(4) \ltimes \bR^8$ solutions with $G=0$ investigated in the previous section. So they
are all locally maximally supersymmetric.

\newsection{$G_2$-invariant normal} \label{G2}

For solutions with a $G_2$-invariant normal $\nu_1$, we take, without loss of generality,
\be
\nu_1 = e_5 + e_{12345} \pm i (e_1 + e_{234})~.
\ee
In particular, $\nu_1$  is invariant under $G_2$ transformations generated by
\be
R^p \big(\Gamma_{1p}-\Gamma_{\bar{1} p} + {1 \over 2} \epsilon_p{}^{\bar{q}_1 \bar{q}_2} \Gamma_{\bar{q}_1
\bar{q}_2} \big) + R^{\bar{p}} \big( \Gamma_{\bar{1} \bar{p}}- \Gamma_{1 \bar{p}}
+{1 \over 2} \epsilon_p{}^{q_1 q_2} \Gamma_{q_1 q_2} \big)~,~~~L^{p \bar{q}} \Gamma_{p \bar{q}}~,
\ee
written in a manifestly $SU(3)\subset G_2$ covariant notation as in \cite{sgiib, sgiib2, iibsyst}, where $L$ is traceless $L^p{}_p=0$ and $p,q=2,3,4$.
Therefore, $L$ generates  $SU(3)$ transformations in the
$2,3,4$ directions.

To choose the second normal $\nu_2$, consider the most general spinor linearly independent from $\nu_1$,
\be
\nu_2 = -x (e_5-e_{12345}) -u^\alpha e_\alpha -{1 \over 2} v^{\alpha \beta} e_{\alpha \beta 5}
-{1 \over 6}w_\alpha \epsilon^{\alpha \beta_1 \beta_2 \beta_3} e_{\beta_1 \beta_2 \beta_3}~,
\ee
where $\a=1,2,3,4$ and similarly for the rest of the indices.

By applying a $SU(3)$ transformation in the directions $2,3,4$ we can, without loss of generality,
set $w_3=w_4=0$, and then apply a $SU(2)$ transformation in the $3,4$ directions to set $u^4=0$.
Hence, we can choose
\bea
\label{secondg2norm}
\nu_2 &=& -x (e_5-e_{12345}) -u^1 e_1 -u^2 e_2 - u^3 e_3 -v^{12} e_{125}-v^{13} e_{135}
-v^{14} e_{145} \nn \\
&-&v^{23} e_{235} - v^{24} e_{245} - v^{34} e_{345} -w_1 e_{234} + w_2 e_{134}~.
\eea

\subsection{Solutions with $G \neq 0$}

Given $\nu_1$ and $\nu_2$, we can compute $G$ using (\ref{gbilin}). Then observe that the basis $(\eta_a)$ in the
space of Killing spinors includes
\be
 \{ e_{35}, e_{45}, e_{1235} \} \ .
\ee
Evaluating the   integrability condition ({\ref{factorint2}}) on these basis elements using CAC,
one finds a number of relations, including
\be
v^{23}=v^{24}=v^{14}=0 \ .
\ee
This simplifies $\nu_2$, and then the basis $(\eta_a)$ of the Killing spinors includes the elements
\be
\{ e_{35}, e_{45}, e_{1235}, e_{14}, e_{13}, e_{23} \}~.
\ee
Applying
({\ref{factorint2}}) to the above basis elements, one finds that
\bea
x=  \mp iu^1~,~~~w_1=-u^1~,
\eea
 ie if one assumes, for example,
that $x \neq \mp iu^1$, then  ({\ref{factorint2}}) implies that
$x=u^1=0$, and similarly for  $w_1=-u^1$. In addition, one obtains the conditions
\be
w_2 = \mp i v^{34}, \quad u^3=v^{13}=0, \quad u^2 = \pm i v^{12}~.
\ee
Next, applying ({\ref{factorint2}}) to the basis elements
\be
 \{ e_{24}, e_{1245}, e_{34} \pm i e_{1345}, e_{12} \pm ie_{25}, 1-e_{1234} \pm i (e_{15}-e_{2345}),
1+e_{1234}\pm i(e_{2345}+e_{15}) \}~,
\ee
one finds  that
\be
v^{34}=u^1=v^{12}=0 \ .
\ee
Combining all the conditions implied by ({\ref{factorint2}}) on the components of $\nu_2$ together,
one concludes that   $\nu_2=0$.  This in turn gives $G=0$ which is a contradiction. Thus there are no
backgrounds with $G\not=0$ in the $G_2$ case which preserve 28 supersymmetries.

\subsection{Solutions with $G=0$}

Consider first the case for which $w_2 \neq 0$ in ({\ref{secondg2norm}}).
By applying gauge transformations to $\nu_1$, $\nu_2$ generated by
$R^1 \Gamma_{-1} + R^{\bar{1}} \Gamma_{- \bar{1}}$ and
$R^3 \Gamma_{-3} + R^{\bar{3}} \Gamma_{- \bar{3}}$,
one can eliminate the $e_{345}$ and the $e_{145}$ terms from
$\nu_2$. However, as these gauge transformations are {\it not} in $G_2$, the
form of $\nu_1$ is not left invariant under their action. Nevertheless,
the simplification to $\nu_2$ produced assists the computation.
After these  transformations, the two normals become
\bea
\nu_1 &=& \alpha e_5 + \beta e_{12345} + \mu e_{135} + \nu e_{245}  \pm i (e_1 + e_{234})
\nn \\
\nu_2 &=& \rho e_5 + \sigma e_{12345} - u^1 e_1 - u^2 e_2 - u^3 e_3
\nn \\
&-&s^{12} e_{125} - s^{13} e_{135}  - s^{23} e_{235} - s^{24} e_{245} - w_1 e_{234} + w_2 e_{134}~.
\eea

To proceed, since $G=0$, the only integrability condition that remains to be satisfied is
\bea
{\cal S}\eta_a\equiv ({1\over 2}T^2+{1\over24}T^4)\eta_a=0~,
\eea
and it is given in detail in (\ref{st2t4}). Evaluating this  integrability condition on the basis elements
\bea
 &{}& \{ e_{35}, e_{45}, e_{1235}, e_{23}, e_{12}, w_2 e_{1345} + u^2 e_{25},
w_2 e_{1245} - u^3 e_{25}, w_2 e_{34} + s^{12} e_{25},
\nn \\
&{}&  w_2 e_{14}+ s^{23} e_{25},
w_2 (e_{15}-e_{2345}) + (u^1-w_1) e_{25}, w_2 1 \pm i \beta w_2 e_{15} + (\sigma \mp i \beta w_1) e_{25},
\nn \\
&{}& w_2 e_{24} \pm i \mu w_2 e_{15} + (-s^{13}\mp i \mu w_1) e_{25},
w_2 e_{13} \pm i \nu w_2 e_{15} + (-s^{24} \mp i \nu w_1) e_{25}  \}
\nn \\
\eea
one finds after  some CAC that
all components of $T^2$ and $T^4$ vanish. Therefore all these solutions are locally maximally
supersymmetric.

Next, consider the case for which $w_2=0$. Then by making a $SU(2)$ rotation in the
$2, 3$ directions, one can set, without loss of generality, $u^3=0$ also.
Suppose first that that $u^2 \neq 0$. By applying a gauge transformation
generated by $R^1 \Gamma_{-1} + R^{\bar{1}} \Gamma_{- \bar{1}}$
to $\nu_1$, $\nu_2$, one can eliminate the $e_{125}$ term from $\nu_2$. Again
the form of $\nu_1$ is altered, because this transformation is
not in $G_2$. Then apply a $SU(2)$ rotation in the $3, 4$ directions to eliminate
the $e_{245}$ term from $\nu_2$. We therefore obtain
\bea
\nu_1 &=& \alpha e_5 + \beta e_{12345} \pm i (e_1 + e_{234})
\nn \\
\nu_2 &=& \rho e_5 + \sigma e_{12345} - u^1 e_1 - u^2 e_2
\nn \\
&-& s^{13} e_{135} - s^{14} e_{145} - s^{23} e_{235} - s^{34} e_{345} - w_1 e_{234}~.
\eea

To proceed, evaluate the ${\cal S}\eta_a=0$ integrability  condition on the basis elements
\bea
 &{}& \{ e_{35}, e_{45}, e_{1235}, e_{25}, e_{1245}, e_{34}, e_{13},
u^2 e_{24} + s^{13} e_{1345}, u^2 e_{23} - s^{14} e_{1345}
\nn \\
&{}& u^2 e_{14} - s^{23} e_{1345}, u^2 e_{12} - s^{34} e_{1345},
u^2 (e_{2345} - e_{15}) +(u^1-w_1) e_{1345},
\nn \\
&{}& u^2 1 \pm i \beta u^2 e_{15} + (-\sigma \pm i \beta w_1) e_{1345}~,
\}
\eea
to find   that all components of $T^2$ and $T^4$ vanish. Thus again
 these solutions are  locally maximally supersymmetric.

Next consider the case for which $w_2=u^2=u^3=0$.
One can then apply a $SU(3)$ transformation to set, without loss of generality
$v^{13}= v^{14}=0$, followed by a $SU(2)$ transformation in the $3,4$ directions
to set $v^{24}=0$. After doing this, we have
\bea
\nu_1 &=& e_5 + e_{12345} \pm i (e_1 + e_{234})
\nn \\
\nu_2 &=& -x (e_5 - e_{12345}) - u^1 e_1 - v^{12} e_{125} - v^{23} e_{235}
- v^{34} e_{345} - w_1 e_{234}~.
\eea
Suppose that $v^{12} \neq 0$. Then apply the integrability condition ${\cal S}\eta_a=0$
to the basis elements

\bea
&{}& \{ e_{35}, e_{45}, e_{1235}, e_{1345},
e_{1245}, e_{24}, e_{23}, e_{13}, e_{25}, v^{12} e_{14} - v^{23} e_{34},
v^{12} e_{12} - v^{34} e_{34},
\nn \\
&{}& v^{12} (e_{2345} - e_{15}) + (u^1-w_1) e_{34},
v^{12} 1 \pm i v^{12} e_{15} +(-x \pm i w_1) e_{34}
\}~,
\eea
 one finds that all components of $T^2$ and $T^4$ vanish, and
so the solutions are again locally maximally supersymmetric.

Finally, consider the case for which $w_2=u^2=u^3=v^{13}=v^{14}=v^{24}=v^{12}=0$.
By making a $SU(2)$ transformation in the $2,4$ directions, one can also set, without
loss of generality, $v^{23}=0$. In such a case,
\bea
\nu_1 &=& e_5 + e_{12345} \pm i (e_1 + e_{234})
\nn \\
\nu_2 &=& -x (e_5 - e_{12345}) - u^1 e_1 - v^{34} e_{345} - w_1 e_{234} \ .
\eea
Next, note that for solutions preserving exactly $N=28$ supersymmetries,
there must be a basis  $(\eta_a)$  for the Killing spinors which contains
the elements
\bea
\label{fincaseg2}
 &&\{ e_{35}, e_{45}, e_{1235}, e_{1345}, e_{1245}, e_{24}, e_{23},
 e_{13}, e_{25}, e_{34}, e_{14},
 \cr
 &&~~~~~~~~~~~~~~z_1 1 + z_2 e_{1234} + z_3 e_{15} + z_4 e_{2345} + z_5 e_{12} \}
\eea
where $z_1$, $z_2$, $z_3$, $z_4$ do not all vanish. Note that this is clearly true if $v^{34} \neq 0$.
In the special case for which $v^{34}=0$, $e_{12}$ can  be taken
as a basis element, but for a $N=28$ solution, two further basis elements must also be found,
and such a solution therefore still has a basis containing the spinors in ({\ref{fincaseg2}}).
On evaluating the integrability condition ${\cal S}\eta_a=0$ on these  basis elements,
one finds that all components of $T^2=T^4=0$. Thus again these solutions are  locally maximally supersymmetric.

Therefore, we have shown that if one of the normals is $G_2$-invariant, then all solutions
with $N=28$ supersymmetries  are locally maximally
supersymmetric.

\newsection{Discrete Quotients}

We have demonstrated that all $N=28$ supersymmetric IIB backgrounds are either locally isometric to that of
\cite{benaroiban} or to a maximally supersymmetric background. The possibility remains that some $N=28$ backgrounds
can be constructed as  discrete quotients of maximally supersymmetric ones. Here, we shall prove  that all
discrete quotients
of maximally supersymmetric backgrounds preserve less than 28 supersymmetries, $N<28$. So there are no $N=28$
backgrounds which arise as discrete quotients of maximally supersymmetric ones.
 To show this,
we shall use the machinery developed in \cite{simon}. This has been applied  both in  M-theory \cite{jose2b} to prove that
there are no $N=31$ quotients   of maximally supersymmetric backgrounds  and in IIB supergravity
  \cite{nm28} to demonstrate a similar result for backgrounds with
$N>28$ supersymmetries.

\subsection{Discrete quotients of Minkowski space }

This computation is similar to that we have performed in \cite{nm28} to search for backgrounds with $N>28$
supersymmetries, so we shall not give an extensive description of the analysis. To find discrete quotients of Minkowski space
which preserve 28 supersymmetries, one has to find an element $\a\in SO(9,1)$ such that its lift $\hat\alpha\in Spin(9,1)$
preserves 28 spinors, ie it acts as the identity on a 28-dimensional subspace of the Weyl representation $\Delta_{{\bf 16}}$ of $Spin(9,1)$.
Up to a conjugation, there are two choices for the lift $\hat\a$. One choice is that $\hat\a$ can be written as
\bea
\hat\a=\exp[{1\over2}(\theta_0\Gamma_{05}+\theta_1\Gamma_{16}+\theta_2\Gamma_{27}+\theta_3\Gamma_{38}+\theta_4\Gamma_{48})+i\psi]~,
\eea
where the additional angle $\psi$ has been added because of the $Spin_c(9,1)$ nature of spinors of IIB supergravity.
Decomposing $\Delta_{\bf 16}$ as
\bea
\Delta_{\bf16}=\sum_{\sigma_0, \dots, \sigma_4=\pm1} W_{\sigma_0\dots\sigma_4}
\eea
using the projectors $\Gamma_i \Gamma_{i+5}$, $i=0,\dots,4$, the lifted element can be written as
\bea
\hat\a(\sigma_0, \dots, \sigma_4)=\exp[{1\over2}(\sigma_0\theta_0+i\sigma_1\theta_1+i\sigma_2\theta_2+i\sigma_3\theta_3+i\sigma_4\theta_4)+i\psi]
\eea
where the chirality condition requires that $\sigma_0\sigma_1\dots \sigma_4=1$.

{}For an element $\hat\a$ to preserve 28 supersymmetries, it has to act as an identity on a 28 dimensional subspace $V$
of $\Delta_{\bf16}$. In particular, there are some $\sigma_0, \dots, \sigma_4$,  such that
\bea
\hat\a(\sigma_0,\sigma_1,\dots, \sigma_4)=1 ~.
\eea
Taking the complex conjugate, one concludes that  $\theta_0=0$. So boosts do not preserve any supersymmetry as expected.
Since $\theta_0=0$, $\hat\a$ is independent of $\sigma_0$. So in what follows we shall explicitly indicate the
dependence of subspaces  $W$ and the map $\hat\a$ on only the
rest of the signs.

To exclude the possibility that some spatial rotations preserve 28 supersymmetries, $\hat\a$ must not act as the identity
on two $W_{\sigma_1\dots\sigma_4}$ subspaces. It is straightforward to observe that whatever the choice
of non-invariant subspaces is, there is always a choice of signs such that
\bea
\hat\a(\sigma_1,\dots, \sigma_4)=\hat\a(-\sigma_1,\dots, -\sigma_4) =1~.
\eea
This in particular implies that $\exp(2i\psi)=1$. Using this, one can show that if for some signs
$\hat\a(\sigma_1,\dots, \sigma_4)=1$, then $\bar{\hat\a}(\sigma_1,\dots, \sigma_4)=\hat\a(-\sigma_1,\dots, -\sigma_4)=1$.
Therefore if the action on $W_{\sigma_1\dots\sigma_4}$ is trivial, so is the action on the conjugate module
$W_{-\sigma_1\dots-\sigma_4}$.
  Thus in order
to preserve precisely 28 supersymmetries, the two non-invariant subspaces should be chosen
to be conjugate to each other.

To proceed since all choices  of the signs are symmetric, without loss of generality, assume that
\bea
\hat\a(+1,+1,+1,+1)={\bar{\hat\a}}(-1,-1,-1,-1)\not=1~.
\eea
To preserve precisely 28 supersymmetries,  for all other choices of signs $\hat\a$ must be the identity.
In particular,
\bea
\hat\a(-1, +1, +1,+1)=\hat\a(1,-1,-1,+1)=1 \ .
\eea
This implies that $\exp(i\theta_4)=1$. This gives
\bea
\exp(i\theta_4) \hat\a(+1, +1,+1, -1)=\hat\a(1,1,1,1)=1
\eea
which is a contradiction. Thus if one assumes that a 28-dimensional subspace of $\Delta_{\bf16}$ is invariant under some
$\hat\a$, then all $\Delta_{\bf16}$ is invariant and so all supersymmetry is preserved. There are no
such quotients with preserve 28 supersymmetries.

The another choice for $\hat\a$  is to take
\bea
\hat\a= \exp\{{1\over2} [(\Gamma_0+\Gamma_5) \Gamma_9+ \theta_1\Gamma_{16}+\theta_2\Gamma_{27}+\theta_3\Gamma_{38}]\}~.
\eea
Decomposing $\Delta_{\bf16}$ using the projector $\Gamma_{05}$, it is easy to see that such quotients preserve at most
16 supersymmetries.

\subsection{Discrete quotients of $AdS_5\times S^5$  }

The isometry group of the $AdS_5\times S^5$ background is  $SO(4,2)\times SO(6)$. To find whether there is a discrete subgroup $D$
of  $SO(4,2)\times SO(6)$ such that $AdS_5\times S^5/D$ preserves 28 supersymmetries, observe that the associated spin group
$Spin(4,2)\times Spin(6)$ acts on $\Delta_{\bf 16}$ as $\Delta^-_{Spin(4,2)}\times \Delta^-_{Spin(6)}$, where
$\Delta^-_{Spin(4,2)}$ and $\Delta^-_{Spin(6)}$ are the chiral representations of $Spin(4,2)$ and $Spin(6)$, respectively.

It is a consequence of the tensor product structure of the representation of $Spin(4,2)\times Spin(6)$ on $\Delta_{\bf 16}$ that
the real dimension of an invariant subspace $V$ of $\hat\a$ is
\bea
{\rm dim} V=2 n m~,~~~1\leq n,m\leq 4~.
\eea
Since $28$ cannot be written as a product in this way, there are no discrete quotients of $AdS_5\times S^5$ which preserve
28 supersymmetries. In fact this argument implies that the largest number of supersymmetries, less than maximal, which can be
preserved by a discrete $AdS_5\times S^5$ quotient\footnote{The addition of the angle $\psi$ due to the $Spin_c$ nature
of IIB spinors does not affect this argument.} is 24.

\subsection{Discrete quotients of Maximally supersymmetric plane wave}

To investigate the existence of discrete quotients of the maximally supersymmetric plane wave
which preserve 28 supersymmetries, we shall follow closely the analysis in \cite{nm28}. In particular,
it has been  shown that the invariance condition for $\hat\a$,  $\hat\a\e=\e$, can be written as
\bea
e^A \epsilon_-&=&\epsilon_-~,
\cr
e^A (\epsilon_++\Gamma_+ \beta \epsilon_-)&=&\epsilon_+~,
\la{actpw}
\eea
where $\Gamma_+\epsilon_+=0$ is the usual light-cone projection. Moreover, one can show that
\bea
\hat\a(\sigma_1, \dots, \sigma_4)\e_-=e^A\epsilon_-=\exp\big[{i\over2} \sum_{i=1}^4 \sigma_i \theta_i+
i\psi\big] \epsilon_-~,~~~\sigma_1\sigma_2\sigma_3\sigma_4=-1~,
\la{signch}
\eea
and
\bea
e^A\, \epsilon_+=\exp\big [-2i \lambda v^- \sigma_1\sigma_2+{i\over2} \sum_{i=1}^4 \sigma_i \theta_i+
i\psi\big]\, \epsilon_+~,~~~\sigma_1\sigma_2\sigma_3\sigma_4=1~.
\la{signchb}
\eea
In particular, the part of $A$ that depends on $v^-$ acts with the identity on $\epsilon_-$. Decomposing
$\Delta_{\bf 16}=V_-\oplus V_+$ using the lightcone projection, it has been shown in \cite{nm28} that there is no
quotient which preserves more than 28 supersymmetries.

To extended the above result to the $N=28$ case, there are two possibilities. Either the discrete group action leaves invariant
a 6-dimensional subspace in $V_-$ and acts as the identity on $V_+$ or vice versa. Consider first the former possibility.
As has already been indicated in (\ref {signch}) and (\ref {signchb}), we have decomposed
both $V_-$ and $V_+$ into eight 1-dimensional complex subspaces $W_{\sigma_1\dots \sigma_4}$ and
$Z_{\sigma_1\dots \sigma_4}$, respectively, labeled by the eight independent choices of signs $\sigma$.
  It is easy to see that
whatever the choice of the 6-dimensional invariant subspace of $V_-$ is, one can show that $e^{2i\psi}=1$.
Using this, one can also show that
if a subspace $W_{\sigma_1\dots \sigma_4}$ is invariant so is the subspace $W_{\bar\sigma_1\dots \bar\sigma_4}$
with $\bar\sigma_i=-\sigma_i$.
Thus for $e^A$ to preserve precisely a 6-dimensional subspace of $V_-$, the non-invariant 2-dimensional complex
subspace of $V_-$
must be as $W_{\sigma_1\dots \sigma_4}\oplus W_{\bar\sigma_1\dots \bar\sigma_4}$ for some choice of $\sigma_i$.
Since the choice of signs is symmetric, without loss of generality, one can choose $W_{+1,+1,+1,-1}\oplus W_{-1,-1,-1,+1}$
as the non-invariant subspace. Solving the condition $e^A=1$ for the remaining choices
of signs, one finds that
\bea
\theta_1=\theta~,~~~\theta_2=\theta+2\pi n_2~,~~~\theta_3=\theta+2\pi n_3~,~~~\theta_4=-\theta+2\pi n_4~,
\la{ang1}
\eea
and
\bea
\pi (n_2+n_3+n_4)+\psi\in 2\pi \bZ~,~~~n_2, n_3, n_4\in \bZ~,
\la{ang2}
\eea
where $\theta$ is an arbitrary angle and $\psi=n\pi$, $n\in \bZ$. So there are transformations which preserve
a six dimensional subspace $I$ of $V_-$.

To preserve precisely 28 supersymmetries, all $V_+$ must be invariant under the action of the discrete group.
For this it is necessary that $I\subset {\rm Ker}\,\beta$ and that $e^A$ acts as the identity on $V_+$.
It is always possible to choose the group action to satisfy the first condition.  So let as focus on the second.
In particular, the invariance of the subspaces $Z_{+1,+1,+1,+1}$ and $Z_{-1,-1,-1,-1}$ imply that
\bea
e^{-2i\lambda v^-+i\theta}= e^{-2i \lambda v^--i\theta}=1~.
\eea
This in turn gives
\bea
2 \lambda v^-=n_0 \pi~,~~~\theta= n_1 \pi~,~~~ n_0+n_1\in 2\bZ~,~~~n_0, n_1\in \bZ~.
\la{xxxx}
\eea
However now notice that for this choice of $\theta$, $W_{+1,+1,+1,-1}$ and so  $W_{-1,-1,-1,+1}$
are also invariant, ie all $V_-$ is preserved.
In such a case, the only option for preserving 28 supersymmetries is that ${\rm dim}_{\bC} {\rm Ker} \,\b=6$. However,
it is easy to see that the dimension of the kernel of $\b$ is either 4 or 8. So such quotients cannot preserve strictly
28 supersymmetries.

Next suppose that the discrete symmetry preserves all $V_-$. In such a case, the angles $\theta_i$
are given as in (\ref{ang1}) and (\ref{ang2}), and $\theta=n_1\pi$, $n_1\in \bZ$. For the quotient to preserve
precisely 28 supersymmetries, one should choose the discrete subgroup that ${\rm dim}_{\bC}{\rm Ker} \b=8$.
 As we have already mentioned
there is always such a choice. We require that $e^A$ leaves invariant a complex 6-dimensional
subspace of $V_+$. In particular, note that
\bea
{\exp\big [ i \sum_{i=1}^4 \sigma_i \theta_i  \big]} =1
\eea
where $\sigma_1 \sigma_2 \sigma_3 \sigma_4=1$, and $\theta_i$ are constrained as above.
It follows that if $Z_{\sigma_1,\sigma_2,\sigma_3,\sigma_4}$ is an invariant subspace, then so is
$Z_{-\sigma_1,-\sigma_2,-\sigma_3,-\sigma_4}$.
Thus for $e^A$ to preserve precisely a 6-dimensional subspace of $V_+$, the non-invariant 2-dimensional complex
subspace of $V_+$
must be as $Z_{\sigma_1\dots \sigma_4}\oplus Z_{-\sigma_1\dots -\sigma_4}$ for some choice of $\sigma_i$.
Since all choices are symmetric, take as the non-invariant subspace
$Z_{+1,+1,+1,+1}\oplus Z_{-1,-1,-1,-1}$. Requiring that $e^A$ leave invariant the $6$-dimensional
subspace complementary to $Z_{+1,+1,+1,+1}\oplus Z_{-1,-1,-1,-1}$ imposes the condition
\bea
e^{-2i \lambda v^- + i \pi n_1} =1 \ .
\eea
However, this condition also implies that $Z_{+1,+1,+1,+1}$ is an invariant subspace, and so
all $V_+$ is invariant. Thus all the supersymmetry is
preserved, and there are
no quotients that preserve strictly 28 supersymmetries.

\newsection{Strings in the plane wave background}

\subsection{Geometry of plane wave}

As we have already mentioned, the plane wave solution (\ref{bena}) is the superposition of two other plane wave solutions, those of
the maximally supersymmetric plane wave of \cite{plwav} and the heterotic
plane wave  preserving 14 supersymmetries\footnote{See \cite{jose2} and \cite{typeI} for a general discussion
of heterotic solutions with more than 8 supersymmetries.}. These two solutions are also recovered in the limits of (\ref{bena})
for which  the parameters $(\ell, h)$ vanish.

We have shown that (\ref{bena})  is a Lorentzian symmetric space and the form fluxes are parallel. In fact
the spacetime is a Lorentzian Lie group because the wave profile is negative definite.
The isometries of the metric are precisely those of the maximally supersymmetric plane wave which have been investigated in \cite{plwav}.
In particular, the algebra of Killing vector fields is $\mathfrak{so}(8)\ltimes \mathfrak{h}(8)$, where  $\mathfrak{h}(8)$ is
the Heisenberg Lie algebra extended by an outer $\mathfrak{u}(1)$ automorphism   which rotates the $8$ positions to the $8$ momenta and commutes
with the central element. However the fluxes are not invariant under the whole group of isometries.
The 5-form flux, as is well known, breaks this group to $(\mathfrak{so}(4)\oplus\mathfrak{so}(4)) \ltimes \mathfrak{h}(8)$.
The additional 3-form flux of (\ref{bena}) breaks the isometry group further to
$(\mathfrak{u}(2)\oplus\mathfrak{u}(2)) \ltimes \mathfrak{h}(8)$ which is the symmetry group of the background. The
$\mathfrak{u}(2)\oplus\mathfrak{u}(2)$ is identified as the subalgebra of $\mathfrak{so}(4)\oplus\mathfrak{so}(4)$ which
in addition preserves a complex structure on the transverse directions to the lightcone.
Moreover observe that in the limit that the 5-form flux vanishes, the symmetry group of the background enhances to
$\mathfrak{u}(4) \ltimes \mathfrak{h}(8)$.

\subsection{String propagation}

The worldvolume dynamics of a string in the (\ref{bena}) background
 is described by a Green-Schwarz action.
  To quantize string theory, one has to gauge fix the kappa symmetry
and rewrite the theory in terms of worldvolume fermions. In this case, this procedure is considerably simplified
because the background is a plane wave and it admits a natural lightcone gauge. In particular, the resulting action
is always
quadratic in the worldvolume fermions \cite{russo}. We shall not carry out this
procedure in detail. Instead, we shall use the close relation that this theory has with the maximally
supersymmetric plane wave
and argue that the bosonic part of the string action is that of a string on a plane wave group manifold
\bea
ds^2&=& 2du dv-(\ell^2 +4h^2) x^2 du^2+dx^2~,
\cr
G&=&-2 \,\ell\, du\wedge  \omega~,~~~~ \omega= (dx^1\wedge dx^6+dx^2\wedge dx^7+dx^3\wedge dx^8+dx^4\wedge dx^9)
\la{gb}
\eea
where we  re-scale $\ell$ to $\ell/\sqrt2$. (The normalization of the fluxes is consistent with that of \cite{btp}.)
In particular,   the 5-form flux does not contribute in the bosonic part of the action apart from
the $h^2$ contribution in the metric. However
 it is expected to contribute in the fermion couplings.

The quantization of strings  on a (\ref{gb}) background is a special case of the models
investigated in \cite{btp}, see also eg \cite{sadri, john, harmak}.
Here we shall carry out
some of the steps in the analysis of \cite{btp} to identify the lightcone string Hamiltonian.
 We shall show that this Hamiltonian
is a linear superposition of infinite many Harmonic oscillator Hamiltonians. To find the frequencies
of these Harmonic oscillators, we
first use a frequency based ansatz to solve the
classical string equations. In particular, one finds that the classical frequencies $\tilde\omega$ satisfy the equation
\bea
\det\big(( \tilde\omega^2- \ell^2-4h^2-4n^2)\delta_{ij}-4 i n\ell \omega_{ij}\big)=0~,~~~n\in \bZ
\eea
which gives
\bea
[(\tilde\omega^2- \ell^2-4h^2-4n^2)^2-16 n^2\ell^2]^4=0~.
\eea
The center of mass mode, $n=0$, has a single frequency
\bea
(\tilde\omega^{(0)})^2= \ell^2+4h^2~.
\eea
For the other modes one has
\bea
(\tilde\omega^{(n)}_{\pm})^2=\pm 4  n \ell+  \ell^2+4h^2+4n^2~.
\eea
Observe that all frequency squares are positive for  $h>0$.

It has been shown in \cite{btp} that the classical frequencies of the string after quantization are identified
with the quantum frequencies of the lightcone string Hamiltonian.  Moreover $\tilde\omega^{(n)}_{\pm}=\tilde\omega^{(-n)}_{\mp}$ and so the $n$ and $-n$ modes pair.
The lightcone Hamiltonian of the string can be written as
\bea
H=\sum_{n\geq 0} H^{(n)}
\eea
where $H^{(n)}$ is the sum of appropriate Harmonic oscillator Hamiltonians.  In particular, one finds that
\bea
H^{(0)}=\sum_{j=1}^8\tilde\omega^{(0)} ({\cal N}_j+{1\over2})~,~~~{\cal N}_j={\rm a}_j^\dagger\,{\rm a}_j~,
\eea
and
\bea
H^{(n)}=\sum_{i=1}^8\tilde\omega_+^{(n)} ({\cal N}_+{}^{(n)}_i+{1\over2})+\sum_{j=1}^8\tilde\omega_-^{(n)} ({\cal N}_-{}^{(n)}_{j}+{1\over2})~,~~~
{\cal N}_\pm{}^{(n)}_j={\rm a}_\pm{}^{(n)}{}^\dagger_j\,\,{\rm a}_\pm{}^{(n)}_j~.
\eea
The operators  ${\rm a}_j^\dagger$, ${\rm a}_\pm{}^{(n)}{}^\dagger_j$ and ${\rm a}_j$, ${\rm a}_\pm{}^{(n)}{}_j$ are creation and annihilation
operators, respectively,  canonically normalized as those of a Harmonic oscillator. So we have shown that the center of mass
Hamiltonian comprises  of 8 Harmonic oscillators with the same frequency and each oscillator mode $n>0$ comprises
of 8 Harmonic oscillators with frequency $\tilde\omega_+^{(n)}$ and 8 Harmonic oscillators with frequency $\tilde\omega_-^{(n)}$.

\newsection{Outlook}

We have shown that the IIB supersymmetric backgrounds with strictly 28 supersymmetries are locally isometric to the
solution of \cite{benaroiban}. Combining this with the classification
of the maximally supersymmetric backgrounds of IIB supergravity in \cite{georgejose} and the results of \cite{n31, nm28} gives a classification
of all supersymmetric backgrounds of IIB supergravity with more than 27 supersymmetries, $N>27$. The conjecture
of \cite{duff} is consistent with our result.
Moreover, we have demonstrated that IIB backgrounds with only 5-form flux
that admit more than 26 supersymmetries, $N>26$, are maximally supersymmetric.

It is not known whether there are IIB solutions which preserve 25, 26 or 27 supersymmetries. However, it is known that there is a
plane wave solution which preserves 24 supersymmetries \cite{benaroiban}. This is again a superposition of the maximally
supersymmetric plane wave with a plane wave solution of the heterotic string which preserves 12 supersymmetries. Since
there is a unique heterotic solution which preserves 12 supersymmetries and  there are no  solutions which preserve 13 supersymmetries, it is tempting
to propose that the IIB $N=24$ solution is unique and there are no IIB solutions  with 25, 26 and 27 supersymmetries. However,
there is no firm evidence for this apart from the analogy with the plane-wave solutions of the heterotic string.

\vskip 0.5cm
\noindent{\bf Acknowledgements}
U.G.~is supported by the Swedish Research Council. J.G.~ is supported by the EPSRC grant, EP/F069774/1.
G.P.~ is partially supported by EPSRC grant, EP/F069774/1, the STFC rolling
grant, PP/C5071745/1, and the EU grant MRTN-2004-512194.
\vskip 0.1cm

\vskip 0.5cm

\setcounter{section}{0}

\appendix{The normals to the Killing spinors}

\subsection{$SU(4)\ltimes \bR^8$}
\subsubsection{Second normal}
A representative for the first $SU(4)\ltimes \bR^8$-invariant normal spinor \cite{n31, nm28} is
\bea
\nu_1=-p e_5- q e_{12345}~,
\eea
where $|p|\not= |q|$.
The infinitesimal generators of the $SU(4)\ltimes \bR^8$  isotropy group are
\be
\label{r8gen}
L^{\alpha \bar{\beta}} \Gamma_{\alpha \bar{\beta}}~,~~~R^{\alpha} \Gamma_{- \alpha} + R^{\bar{\alpha}}
\Gamma_{- \bar{\alpha}}~,~~~\a,\b=1,2,3,4~,
\ee
where $L\in \mathfrak{su}(4)$, ie $L^{\alpha \bar{\beta}} \delta_{\alpha \bar{\beta}}=0$, and
$R^{\alpha} = (R^{\bar{\alpha}})^*$.

A basis in the anti-chiral $Spin(9,1)$ representation $\Delta^-_{\bf 16}$ can be chosen as
\bea
e_5~,~~~e_{12345}~,~~~e_\mu~~~~e_{\mu\nu\rho}~,~~~e_{\mu\nu5}~ \ .
\eea
$\Delta^-_{\bf 16}$ is decomposed under $SU(4)$ as  ${\bf 16}={\bf 1}\oplus {\bf 1}\oplus {\bf 4}\oplus
\bar{\bf 4}\oplus {\bf 6}$.
To find a representative for the second normal, consider the following formulae,
\bea
\big(R^{\alpha} \Gamma_{- \alpha} + R^{\bar{\alpha}}
\Gamma_{- \bar{\alpha}}\big) e_{\mu} &=& 2 R^{\alpha}
e_{\alpha \mu 5} +2 R_\mu e_5~,
\nn \\
\big(R^{\alpha} \Gamma_{- \alpha} + R^{\bar{\alpha}}
\Gamma_{- \bar{\alpha}}\big)  e_{\mu \nu 5} &=&0~,
\nn \\
\big(R^{\alpha} \Gamma_{- \alpha} + R^{\bar{\alpha}}
\Gamma_{- \bar{\alpha}}\big) e_{\mu \nu \rho}
&=& 2 R^\alpha \epsilon_{\alpha \mu \nu \rho} e_{12345}
+6 R_{[\mu} e_{\nu \rho] 5}~,
\eea
and
\bea
L^{\alpha \bar{\sigma}} \Gamma_{\alpha \bar{\sigma}} e_{\mu}
&=& 2 L^\alpha{}_\mu e_{\alpha}
\nn \\
L^{\alpha \bar{\sigma}} \Gamma_{\alpha \bar{\sigma}} e_{\mu \nu 5}
&=& -4 L^\alpha{}_{[\mu} e_{\nu] \alpha 5}~,
\nn \\
L^{\alpha \bar{\sigma}} \Gamma_{\alpha \bar{\sigma}}
\epsilon^{\gamma \beta_1 \beta_2 \beta_3} e_{\beta_1 \beta_2 \beta_3}
 &=& -2 L^\gamma{}_\rho  \epsilon^{\rho \beta_1 \beta_2 \beta_3} e_{\beta_1 \beta_2 \beta_3}~.
\eea

It is also useful to consider the
gauge transformations generated by $\Gamma_{+-}$ and
$i \delta^{\alpha \bar{\beta}} \Gamma_{\alpha \bar{\beta}}$ which
act on spinors as
\bea
e^{f_1 \Gamma_{+-} + i f_2 \delta^{\alpha \bar{\beta}} \Gamma_{\alpha \bar{\beta}}} e_5 &=& e^{-f_1-4if_2} e_5~,
\nn \\
e^{f_1 \Gamma_{+-} + i f_2 \delta^{\alpha \bar{\beta}} \Gamma_{\alpha \bar{\beta}}} e_{12345} &=& e^{-f_1+4if_2}
e_{12345}~,
\nn \\
e^{f_1 \Gamma_{+-} + i f_2 \delta^{\alpha \bar{\beta}} \Gamma_{\alpha \bar{\beta}}} e_{\mu} &=& e^{f_1-2if_2}
e_{\mu}~,
\nn \\
e^{f_1 \Gamma_{+-} + i f_2 \delta^{\alpha \bar{\beta}} \Gamma_{\alpha \bar{\beta}}} e_{\mu \nu 5} &=& e^{-f_1}
e_{\mu \nu 5}~,
\nn \\
e^{f_1 \Gamma_{+-} + i f_2 \delta^{\alpha \bar{\beta}} \Gamma_{\alpha \bar{\beta}}} e_{\mu \nu \rho}
&=& e^{f_1+2if_2} e_{\mu \nu \rho }~.
\eea
Although these transformations do not leave $e_5$ and $e_{12345}$ invariant, they do leave
the plane  spanned of $e_5$ and $e_{12345}$ invariant. So they are generators of the $\Sigma$ group \cite{typeI}.

Now suppose that
\be
\nu_2 = -X e_5 -Y e_{12345} - u^\alpha e_{\alpha} - {1 \over 2} v^{\alpha \beta} e_{\alpha \beta 5}
- {1 \over 6} w_\alpha \epsilon^{\alpha \beta_1 \beta_2 \beta_3} e_{\beta_1 \beta_2 \beta_3}~,
\ee
is the second normal spinor.

Using a $SU(4)$ transformation, we can without loss of generality
set $w_2=w_3=w_4=0$, with $w_1=w$. The isotropy group is $SU(3)$.
 By applying a $SU(3)$ transformation
in the $2,3,4$ directions, one can without loss of generality also set
$u^3=u^4=0$. So we have
\be
\nu_2 = -X e_5 - Y e_{12345} - u^1 e_{1} - u^2 e_{2} - {1 \over 2} v^{\alpha \beta} e_{\alpha \beta 5} -w e_{234}~.
\ee

Next apply a $\bR^8$ gauge transformation generated by $R^\alpha \Gamma_{- \alpha}+
R^{\bar{\alpha}} \Gamma_{- \bar{\alpha}}$, which maps
\bea
\nu_2 \rightarrow \nu'_2 &=& \big(-X-2R_1 u^1-2R_2 u^2\big) e_5
+ \big(-Y-2w R^1 \big) e_{12345} -u^1 e_1 -u^2 e_2 -w e_{234}
\nn \\
&+& \big(-v^{1p}+2 u^1 R^p -2 u^p R^1 \big) e_{1p5}
+ \big(-{1 \over 2} v^{p_1 p_2} +2 u^{[p_1} R^{p_2]}-w \epsilon^{q p_1 p_2} R_q \big) e_{p_1 p_2 5}~,
\nn \\
\eea
for $p,q=2,3,4$.

First consider the case for which $|u^1|^2+|u^2|^2 \neq 0$. Then one can choose
$R^1, R^2, R^3, R^4$ in order to eliminate the $e_5$ and $e_{1p5}$ terms ($p=2,3,4$),
giving
\be
\nu_2 = -y e_{12345} -u^1 e_1 -u^2 e_2 -w e_{234} -{1 \over 2} (v')^{p_1 p_2} e_{p_1 p_2 5}~.
\ee
In addition by applying a $SU(2)$ transformation in the $3,4$ directions, one can remove the
$e_{245}$ term, to leave
\be
\label{normal2a}
\nu_2 = -y e_{12345} -u^1 e_1 -u^2 e_2 -w e_{234} -c_3 e_{235} - c_4 e_{345}~.
\ee

Next consider the case for which $u^1=u^2=0$. In this case applying the $\bR^8$ transformation gives
\be
\nu'_2 = -X e_5 + \big(-Y-2w R^1 \big) e_{12345} -w e_{234}-v^{1p} e_{1p5}
+ \big(-{1 \over 2} v^{p_1 p_2}-w \epsilon^{q p_1 p_2} R_q\big) e_{p_1 p_2 5}~.
\ee
If $w \neq 0$, then one can choose $R^1, R^2, R^3, R^4$ in order to eliminate the
$e_{12345}, e_{235}, e_{245}, e_{345}$ terms, giving
\be
\nu_2 = -X e_5 -w e_{234} -v^{1p} e_{1p5}~.
\ee
Then, applying a $SU(3)$ transformation in the $2,3,4$ directions, the $e_{125}$ and $e_{135}$ terms
can also be removed to give
\be
\label{normal2b}
\nu_2 = -x e_5 -w e_{234} - c_3 e_{145}~.
\ee
However, note that this ({\ref{normal2b}}) is gauge equivalent
to a special case of ({\ref{normal2a}}). The gauge transformation
used to relate the two $\nu_2$ is $\Gamma_{1234}= e^{{\pi \over 2}(\Gamma_{12}+\Gamma_{34})}$
(here the indices are in the {\it real} basis). Furthermore, this gauge transformation also
preserves the span of $e_5$ and $e_{12345}$. Hence we can discard the case when $w \neq 0$.

The remaining case therefore has $u^1=u^2=w=0$. Then
\be
\nu_2 = -X e_5 -Y e_{12345} -{1 \over 2} v^{\alpha \beta} e_{\alpha \beta 5}~.
\ee
By applying a $SU(4)$ transformation, as set out in Appendix A of \cite{mtheor},
one can work in  a gauge for which
\be
\nu_2 = -x e_5 -y e_{12345} -c_1 e_{145} -c_2 e_{235}~.
\ee

\subsubsection{Null planes}

 We have already demonstrated above how to choose the two normal spinors $(\nu_1, \nu_2)$ up to
 $SU(4)\ltimes\bR^8$ transformations. The choice of the second spinor can be simplified further.
For this, observe that if a direction in the space of the two normals $(\nu_1, \nu_2)$
is associated with a time-like vector bilinear, then the corresponding background is a special case of those that will be investigated in
section \ref{G2}. Hence, it suffices to consider only those  $SU(4)\ltimes\bR^8$ cases
 for which all linear combinations
of the two normals $\nu_1$ and $\nu_2$ are associated with null 1-form bilinears.

As we have shown above, there are two choices for the second normal given by
\be
\label{normal2ax}
\nu_2 = -y e_{12345} -u^1 e_1 -u^2 e_2 -w e_{234} -c_3 e_{235} - c_4 e_{345}~,
\ee
with $|u^1|^2+|u^2|^2 \neq 0$,
and
\be
\label{normal2bxb}
\nu_2 = -x e_5 -y e_{12345} -c_1 e_{145} -c_2 e_{235}~.
\ee

For $\nu_2$ given in (\ref{normal2ax}), to impose the condition that the 1-form bilinear
\be
\kappa_M = B(\nu_2 + k \nu_1 , \Gamma_M C (\nu_2+k \nu_1)^*)
\ee
is null for all $k$, we first compute $\kappa^2$ for $k=0$ to find
\be
\kappa^2 = -4 \big((|u^1|^2+|u^2|^2)(|y|^2+|c_4|^2)+|c_3|^2 |u^1|^2 \big)~.
\ee
This vanishes provided we take $y=c_4=0$, and either $c_3=0$ or $u^1=0$.

If $c_3=c_4=y=0$ then the norm of $\kappa$, when $k=1$, is given by
$-4|u^2|^2|q|^2 -4 |\bar{w} p + u^1 \bar{q}|^2$.
Then, either $q=0$ or $w=0$, and one can make a $SU(4)$ gauge transformation to set
\be
\nu_1 = e_5, \qquad \nu_2 = c e^1
\ee
for $c \neq 0$, or $q \neq 0$, $u^2 = 0$, $u^1 \neq 0$ and $\bar{w} p + u^1 \bar{q} =0$.
Thus one finds
\be
\nu_2 = -y e_{12345} - u^1 e_1 - w e_{234} \ .
\ee
Note that in this case, $w \neq 0$.

If, however, $c_4=y=u^1=0$, then the norm of $\kappa$, when $k=1$, is given by
$-4 |u^2|^2 |q|^2 -4 |w|^2 |p|^2$. Requiring this to vanish forces $q=w=0$ and so
\be
\nu_2 = - u^2 e_2 - c_3 e_{235} \ .
\ee
However, this normal is gauge equivalent to $\nu_2 = c e_1$ under an appropriately chosen
$SU(4) \ltimes \bR^8$ gauge transformation.

Hence, requiring that all linear combinations of $\nu_1, \nu_2$  generate null 1-forms reduces
 ({\ref{normal2ax}}) to two simpler sub-cases, with either
 \be
 \label{norm2ax1}
\nu_1 = e_5, \qquad \nu_2 = c e^1 \qquad (c \neq 0)
\ee
or
 \be
 \label{norm2ax2}
 \nu_1 = - p e_5 -q e_{12345} , \qquad \nu_2 = -y e_{12345} - u^1 e_1 - w e_{234}, \
\ee
with $\bar{w} p + u^1 \bar{q} =0$ and non-vanishing $p,w,q, u^1$.

It should be noted that for the case of $\nu_2$ given in ({\ref{normal2bxb}}), all linear
combinations of $\nu_1, \nu_2$ automatically generate null 1-forms, with no
additional constraints on the coefficients in the normals.

\subsection{$Spin(7)\ltimes \bR^8$}

The $Spin(7)\ltimes\bR^8$ case is a special case of the $SU(4)\ltimes \bR^8$ one. An inspection
of section \ref{su4} for $G \neq 0$ reveals that, for all cases that $Spin(7)\ltimes\bR^8$ arises as a special case of  $SU(4)\ltimes \bR^8$,
the normal spinors of the former can be chosen as
\be
\nu_1 = e_5+ e_{12345}~,~~~\nu_2 = x(e_5-e_{12345})+c(e_{145}+e_{235})~,
\ee
where $x,c$ are complex functions. The choice of normals can be further simplified. For this,
choose a basis in the space of Killing spinors normal to $(\nu_1, \nu_2)$ as
\bea
(\eta_a) = \{e_{15}, e_{25}, e_{35}, e_{45}, e_{1235}, e_{1245}, e_{1345}, e_{2345},
e_{12}, e_{13}, e_{24}, \\ \nn e_{34}, e_{23}-e_{14}, c(1-e_{1234})-x(e_{23}+e_{14}) \}~.
\eea
Substituting this into the integrability condition
({\ref{factorint2}}), one finds that   $ x c^*$ is a {\it real} valued function. Hence, without loss of generality, we can set
\be
\nu_2 = e^{i \theta} ( \rho_1 (e_5-e_{12345}) + \rho_2 (e_{145}+e_{235}))~,
\ee
where $\theta, \rho_1, \rho_2 $ are {\it real} functions.
Using the gauge transformation $e^{\phi(\Gamma_{14}+\Gamma_{23})}$, where the gamma matrices are in the real basis and so $\phi$ is real,
 one can  set $\rho_2=0$.
Therefore, the two normal spinors can be chosen as
\be
\nu_1 =e_5+ e_{12345}, \quad \nu_2 = c(e_5-e_{12345}), \qquad (c \neq 0)~.
\ee

\appendix{Gravitino Integrability condition}
The integrability condition of the KSE is
{  \bea
[{\cal D}_N, {\cal D}_M]\e\equiv {\cal R}_{NM}\e= 2{\cal S}\e-2{\cal T}C\e^*
\eea
where
{\small\bea
{\cal S}&=&{1\over 8}R_{NM}{}^{L_1L_2}\Gamma_{L_1L_2}
+
{i\over 48}\Gamma^{L_1\dots L_4}D_{[N}F_{M]L_1\dots L_4} \nn\\
&&+{1\over 24}(
-\Gamma^{L_1L_2}F_{[N|L_1}{}^{Q_1Q_2Q_3}F_{|M]L_2 Q_1Q_2Q_3}
+{1\over 2}\Gamma^{L_1\dots L_4}F_{NML_1}{}^{Q_1Q_2}F_{L_2L_3L_4Q_1Q_2}\nn\\
&&~~~~~~~~~~~~~~~~~~~~~~~~+
{1\over 2}\Gamma_{[N}{}^{L_1L_2 L_3}F_{M]L_1}{}^{Q_1Q_2Q_3}
F_{L_2L_3Q_1Q_2Q_3}) \nn\\
&&+{1\over 32}(
-{1\over2}G_{[N}{}^{L_1L_2}G^\star_{M]L_1L_2}
+{1\over48}\Gamma_{NM}G^{L_1L_2L_3}G^\star_{L_1L_2L_3}\nn\\
&&~~~~~~-{1\over4}\Gamma_{[N}{}^{L_1}G_{M]}{}^{L_2L_3}G^\star_{L_1L_2L_3}
+{1\over8}\Gamma_{[N|}{}^{Q}G_{Q}{}^{L_1L_2}G^\star_{|M]L_1L_2}\nn\\
&&~~~~~~+{3\over16}\Gamma^{L_1L_2}G_{NM}{}^{L_3}G^\star_{L_1L_2L_3}
-\Gamma^{L_1L_2}G_{[N|L_1}{}^{Q}G^\star_{|M]L_2Q}\nn\\
&&~~~~~~-{3\over16}\Gamma^{L_1L_2}G_{L_1L_2}{}^{Q}G^\star_{NMQ}
+{1\over16}\Gamma_{NM}{}^{L_1L_2}G_{L_1}{}^{Q_1Q_2}G^\star_{L_2Q_1Q_2}
\nn\\
&&~~~~~~-{1\over16}\Gamma^{L_1\dots L_4}G_{L_1L_2L_3}G^\star_{NML_4}
+{1\over8}\Gamma_{[N|}{}^{L_1L_2L_3}G_{L_1L_2}{}^{Q}G^\star_{|M]L_3Q}\nn\\
&&~~~~~~+{1\over4}\Gamma^{L_1\dots L_4}G_{[N|L_1L_2}G^\star_{|M]L_3L_4}
+{1\over16}\Gamma^{L_1\dots L_4}G_{NML_1}G^\star_{L_2L_3L_4}\nn\\
&&~~~~~~+{1\over4}\Gamma_{[N|}{}^{L_1L_2L_3}G_{|M]L_1}{}^{Q}G^\star_{L_2L_3Q}
+{1\over24}\Gamma_{[N|}{}^{L_1\dots L_5}G_{|M]L_1L_2}G^\star_{L_3L_4L_5}\nn\\
&&~~~~~~-{1\over48}\Gamma_{[N|}{}^{L_1\dots L_5}
G_{L_1L_2L_3}G^\star_{|M]L_4L_5}
-{1\over32}\Gamma_{NM}{}^{L_1\dots L_4}G_{L_1L_2}{}^{Q}G^\star_{L_3L_4Q}\nn\\
&&~~~~~~-{1\over288}\Gamma_{NM}{}^{L_1\dots L_6}
G_{L_1L_2L_3}G^\star_{L_4L_5L_6})~,
\label{aurelius}
\eea}}

and

{\small\bea
{\cal T}&=&-{1\over96}(
\Gamma_{[N}{}^{L_1L_2L_3}D_{M]}G_{L_1L_2L_3}
+9\Gamma^{L_1L_2}D_{[N}G_{M]L_1L_2})\nn\\
&&+{i\over 32}(
{1\over3}F_{NM}{}^{L_1L_2L_3}G_{L_1L_2L_3}
+\Gamma^{L_1L_2}F_{[N|L_1L_2}{}^{Q_1Q_2}G_{|M]Q_1Q_2} \nn\\
&&~~~~~~+{1\over3}\Gamma_{[N}{}^{Q}F_{M]Q}{}^{L_1L_2L_3}G_{L_1L_2L_3}
-{1\over2}\Gamma^{L_1\dots L_4}F_{NML_1L_2}{}^{Q}G_{L_3L_4Q}\nn\\
&&~~~~~~+{1\over2}\Gamma_{[N}{}^{L_1L_2L_3}F_{M]L_1L_2}{}^{Q_1Q_2}G_{L_3Q_1Q_2}
+{1\over4}\Gamma^{L_1\dots L_4}F_{L_1\dots L_4}{}^{Q}G_{NMQ}\nn\\
&&~~~~~~-{1\over2}\Gamma_{[N|}{}^{L_1L_2L_3}F_{L_1L_2L_3}{}^{Q_1Q_2}
G_{|M]Q_1Q_2})~.
\label{plotinus}
\eea}

\appendix{Integrability condition }

In this appendix, we shall solve the integrability condition (\ref{simpspin7int1})
\be
{(\hat{T}_{NM}})_{L_1 L_2 L_3 L_4} \Gamma^{L_1 L_2 L_3 L_4} \eta_a =0~,
\ee
for $(\eta_a)$ given in (\ref{kspin7}) to show that $\nabla F=0$,
where
\be
({\hat{T}}_{NM})_{L_1 L_2 L_3 L_4} =
D_{[N} F_{M] L_1 L_2 L_3 L_4}~.
\ee
A straightforward but tedious calculation implies that all components
of ${\hat{T}}$ are constrained to vanish, except for
${(\hat{T}_{NM})}_{\alpha_1 \alpha_2 \alpha_3 \alpha_4}$
and ${(\hat{T}_{NM})}_{+ \alpha_1 \alpha_2 \alpha_3}$
(and their complex conjugates)
where $\alpha, \beta$ denote holomorphic indices
in the standard holomorphic light-cone basis.
In fact, these components also vanish. To see this, we make use of the conditions (\ref{feqcond}) on $T^4$. These imply in particular that

\be
\label{icx1}
({\hat{T}}_{[M N})_{L_1 L_2 L_3 L_4]} = 0 \ ,
\ee
\be
\label{icx2}
({\hat{T}}_{L_1 (M})_{N) L_2 L_3 L_4}=({\hat{T}}_{[L_1 |(M})_{N)| L_2 L_3 L_4]} \ ,
\ee
and
\be
\label{icx3}
({\hat{T}}_{M [N_1})_{N_2 N_3 N_4 N_5]}
= -{1 \over 5!} \epsilon_{N_1 N_2 N_3 N_4 N_5}{}^{M_1 M_2 M_3 M_4 M_5}
({\hat{T}}_{M [M_1})_{M_2 M_3 M_4 M_5]} \ .
\ee
Furthermore, as $F= e^+ \wedge \Phi$, and $c e^+$ is covariantly constant,
it follows that
\be
\label{vanTT}
({\hat{T}}_{\tilde{M} \tilde{N}})_{\tilde{L}_1
\tilde{L}_2 \tilde{L}_3 \tilde{L}_4}=0
\ee
where $\tilde{N}$ and the other similar indices take all values except for $``+"$.
This last property implies that
\be
({\hat{T}}_{\tilde{M} [\bar{\beta}})_{\mu_1 \mu_2 \mu_3 \mu_4]}=0 \ .
\ee
The self-duality of ${\hat{T}}$ on the anti-symmetrized indices implies
\be
({\hat{T}}_{\tilde{M} [+})_{- \lambda_1 \lambda_2 \lambda_3]} =0
\ee
and hence
\be
({\hat{T}}_{\tilde{M} -})_{+ \lambda_1 \lambda_2 \lambda_3} =0
\ee
for ${\tilde{M}} =  \alpha, \bar{\alpha}$.
Next, observe that ({\ref{icx2}}) implies that
\be
({\hat{T}}_{+ \bar{\alpha}})_{+ \alpha_1 \alpha_2 \alpha_3}=0, \qquad
({\hat{T}}_{+ -})_{+ \alpha_1 \alpha_2 \alpha_3}=0
\ee
on symmetrizing on $\bar{\alpha}, \alpha_1$ and $-, \alpha_1$ respectively.
Furthermore, ({\ref{icx2}}) also implies that
\be
({\hat{T}}_{\bar{\alpha} \bar{\beta}})_{+ \alpha_1 \alpha_2 \alpha_3}=0
\ee
on symmetrizing appropriately in $\bar{\alpha}, \alpha_1, \bar{\beta}, \alpha_2$.
In addition, the self-duality condition ({\ref{icx3}}) implies that
\be
\label{auxsd1}
({\hat{T}}_{M[+})_{\alpha_1 \alpha_2 \alpha_3 \alpha_4]}=0 \ .
\ee
On setting $M=+$ in ({\ref{auxsd1}})
one finds
\be
({\hat{T}}_{+ [\alpha_1})_{|+|\alpha_2 \alpha_3 \alpha_4]}=0 \ .
\ee
However, ({\ref{icx2}}) implies that
$({\hat{T}}_{+ \alpha_1})_{+\alpha_2 \alpha_3 \alpha_4}$
 is totally antisymmetric in $\alpha_1$, $\alpha_2$, $\alpha_3$, $\alpha_4$, and hence
\be
({\hat{T}}_{+ \alpha_1})_{+\alpha_2 \alpha_3 \alpha_4}=0 \ .
\ee
On setting $M={\bar{\beta}}$ in ({\ref{auxsd1}}), one finds the condition
\be
({\hat{T}}_{\bar{\beta} +})_{\alpha_1 \alpha_2 \alpha_3 \alpha_4}
-4 ({\hat{T}}_{\bar{\beta} [\alpha_1})_{|+|\alpha_2 \alpha_3 \alpha_4]}=0 \ .
\ee
However, ({\ref{icx2}}) implies that $({\hat{T}}_{\bar{\beta} \alpha_1})_{+\alpha_2 \alpha_3 \alpha_4}$
 is totally antisymmetric in $\alpha_1$, $\alpha_2$, $\alpha_3$, $\alpha_4$, and furthermore that
 \be
 ({\hat{T}}_{\bar{\beta} +})_{\alpha_1 \alpha_2 \alpha_3 \alpha_4} = - ({\hat{T}}_{\bar{\beta} \alpha_1})_{+ \alpha_2 \alpha_3 \alpha_4} \ .
 \ee
It follows that
\be
({\hat{T}}_{\bar{\beta} +})_{\alpha_1 \alpha_2 \alpha_3 \alpha_4}  =
 ({\hat{T}}_{\bar{\beta} \alpha_1})_{+ \alpha_2 \alpha_3 \alpha_4}  =0 \ .
\ee

On setting $M=\alpha$ in ({\ref{auxsd1}}), and noting that for a non-zero expression one can take
without loss of generality
$\alpha_1, \alpha_2, \alpha_3, \alpha_4$ to be distinct, with $\alpha=\alpha_1$, it is straightforward
to show that
\be
({\hat{T}}_{\alpha+ })_{ \alpha_1\alpha_2 \alpha_3 \alpha_4} =0 \ ,
\ee
where ({\ref{icx2}}) has also been used.
Next, consider $({\hat{T}}_{\alpha_1 \alpha_2 })_{+ \beta_1\beta_2 \beta_3}$;
without loss of generality one can take $\alpha_1=\beta_1$, then on using
({\ref{icx2}}) to symmetrize on the $+, \alpha_2$ indices, one finds
\be
({\hat{T}}_{\alpha_1 \alpha_2 })_{+ \beta_1\beta_2 \beta_3}=0 \ .
\ee
Hence, we have shown $({\hat{T}}_{MN})_{+ \alpha_1 \alpha_2 \alpha_2}=0$ for all $M,N$.

To proceed, note that ({\ref{vanTT}}) implies that
\bea
({\hat{T}}_{\alpha_1 \alpha_2})_{\beta_1 \beta_2 \beta_3 \beta_4} &=& 0, \qquad
({\hat{T}}_{\alpha \bar{\beta}})_{\beta_1 \beta_2 \beta_3 \beta_4}=0, \qquad
({\hat{T}}_{\bar{\alpha}_1 \bar{\alpha}_2})_{\beta_1 \beta_2 \beta_3 \beta_4}=0,  \\ \nn
({\hat{T}}_{- \alpha})_{\beta_1 \beta_2 \beta_3 \beta_4} &=& 0, \qquad
({\hat{T}}_{- \bar{\alpha}})_{\beta_1 \beta_2 \beta_3 \beta_4} =0 \ ,
\eea
and on setting $M=-$ in   ({\ref{auxsd1}}) one also finds
\be
({\hat{T}}_{- +})_{\beta_1 \beta_2 \beta_3 \beta_4} \ .
\ee

Hence $({\hat{T}}_{MN})_{\alpha_1 \alpha_2 \alpha_3 \alpha_4}=0$ for all $M,N$;
so $\hat T=0$.
In turn, this and  the Bianchi identity for $F$ imply  that
 $\nabla F=0$ as in the case of the maximally supersymmetric backgrounds in \cite{georgejose}.

\end{document}